\documentclass[hidelinks,AMA, STIX1COL ]{WileyNJD-v2}

\usepackage{graphicx}
\usepackage{amsmath}
\usepackage{subcaption}
\usepackage{bm}
\usepackage{fancyvrb}

\DeclareMathOperator{\expit}{expit}
\DeclareMathOperator{\I}{I}

\articletype{}%

\begin{document}

\title{A Practical Introduction to Bayesian Estimation of Causal Effects: Parametric and Nonparametric Approaches\protect}

\author[1]{Arman Oganisian*}

\author[2]{Jason A. Roy}

\authormark{Oganisian \textsc{et al}}

\address[1]{\orgdiv{Department of Biostatistics, Epidemiology, and Informatics}, \orgname{University of Pennsylvania}, \orgaddress{\state{Pennsylvania}, \country{USA}}}

\address[2]{\orgdiv{Department of Biostatistics and Epidemiology}, \orgname{Rutgers University}, \orgaddress{\state{New Jersey}, \country{USA}}}

\corres{*\email{aoganisi@upenn.edu}}

\presentaddress{Room 108, Blockley Hall, 423 Guardian Drive, Philadelphia, PA, USA 19104.}

\abstract[Summary]{Substantial advances in Bayesian methods for causal inference have been made in recent years. We provide an introduction to Bayesian inference for causal effects for practicing statisticians who have some familiarity with Bayesian models and would like an overview of what it can add to causal estimation in practical settings. In the paper, we demonstrate how priors can induce shrinkage and sparsity in parametric models and be used to perform probabilistic sensitivity analyses around causal assumptions. We provide an overview of nonparametric Bayesian estimation and survey their applications in the causal inference literature. Inference in the point-treatment and time-varying treatment settings are considered. For the latter, we explore both static and dynamic treatment regimes. Throughout, we illustrate implementation using off-the-shelf open source software. We hope to leave the reader with implementation-level knowledge of Bayesian causal inference using both parametric and nonparametric models. All synthetic examples and code used in the paper are publicly available on a companion GitHub repository.}



\maketitle

\section{Introduction}
Causal inference is broadly concerned with estimating parameters governing the causal mechanisms between an intervention or treatment of interest and an outcome. These causal parameters can differ substantially from associational ones. Causal inference provides a framework for 1) constructing different estimands that have explicitly causal, rather than associational, interpretations 2) formulating the assumptions under which we can estimate these using observed data, 3) devising sensitivity analyses around violations of these assumptions, and 4) making inferences about these causal estimands via statistical modeling. These are just some of the many contributions of the causal inference literature, which we will touch on throughout this paper.

Bayesian modeling in causal inference has been growing in popularity. There are perhaps several reasons for this phenomenon. First, Bayesian modeling yields full posterior inference for any function of model parameters. For instance, point and interval estimates can be easily constructed for causal risk ratios, odds ratios, and risk differences by post-processing a single set of posterior draws from logistic regression model. Another advantage is the use of priors to induce shrinkage and sparsity in causal models - yielding more regularized causal effect estimates. We show that these can be more satisfying than the ad-hoc alternatives often employed. Priors can also be used to conduct probabilistic sensitivity analyses around violations of key causal identification assumptions. Finally, the Bayesian literature consists of a large suite of nonparametric models that can be readily applied to causal modeling. These nonparametric approaches are appealing because, unlike many classical machine learning algorithms, they allow for posterior uncertainty estimation as well as robust point estimates.

We begin with an overview of the causal identification and the Bayesian linear model before moving to confounder adjustment via standardization in the point-treatment setting. Here we highlight how priors can be used to induce shrinkage in a causal dose effect curve and partially pool conditional average treatment effect estimates across sparsely populated subgroups. Partial pooling shrinks the heterogeneous effects across subgroups towards an overall homogenous effect in the absence of data. We introduce the Bayesian bootstrap as a method for performing standardization. Next, we move to the time-varying treatment and confounding setting where we discuss Bayesian $g$-computation with priors that promote sparsity. Causal inference in these settings requires estimation of a large number of nuisance parameters. Priors that regularize these estimates by encouraging sparsity can be an attractive alternative to common modeling assumptions, which can be quite strict. Estimation for both static and dynamic treatment regimes are discussed. We then turn to using priors for causal sensitivity analyses. These follow from expressing violations of causal assumptions in terms of non-identifiable parameters, then conveying uncertainty about the magnitude/direction of the violation via priors on these parameters. We end with a discussion of Bayesian nonparametric causal estimation. We discuss Dirichlet process priors, Bayesian Additive Regression Trees, Gaussian processes and survey their applications to causal problems. Throughout, we present several pedagogical examples using publicly available synthetic data. We hope to demonstrate how readily these models can be implemented using existing software. A companion GitHub\footnote[3]{\url{https://github.com/stablemarkets/intro_bayesian_causal}} repository contains all relevant implementation code that reproduce the results in this paper.

\section{Ingredients of Bayesian Causal Inference} \label{sc:ingredients}

\subsection{Causal Estimands and Identification Assumptions}
In order to make causal inferences, we first need to define and identify the causal estimand. After doing so, we will turn to the statistical problem of actually making inferences about this estimand. Consider estimation of the causal effect of a binary treatment assignment $A\in\{0,1\}$ on some observed scalar outcome $Y$. In this paper, we formulate estimands in terms of potential outcome $Y^a$ \cite{Rubi:1974}. This represents the outcome that would have been observed had a subject received treatment $A=a$. For subjects receiving treatment $A=a$, we never observe their counterfactual outcome, $Y^{1-a}$. If we did observe both potential outcomes, we could estimate each subject's individual-level effects by taking the difference $Y^1 - Y^0$. This is the difference in outcome had the subject taken treatment 1 versus 0. We could also estimate a population-level \textit{average} treatment effect (ATE), $\Psi = E[Y^1] - E[Y^0]$, directly by simply taking the sample average of the difference, $Y^1 - Y^0$, across all subjects. The ATE is interpreted as the \textit{average} difference in the outcome had everyone in the target population received treatment $A=1$ rather than $A=0$. In absence of the counterfactual, we cannot estimate the individual-level effects and can only estimate $\Psi$ under certain \textit{identification assumptions} ($IA$s).

To understand the role of these assumptions, it is helpful to consider the data we actually observe. Along with $Y$ (note the lack of superscript) and $A$, we observe a vector of confounders $L$ - these are variables, measured pre-treatment, that impact both treatment assignment and outcome. Thus, we could estimate the conditional outcome mean, or \textit{regression},  $E[Y \mid A, L]$ directly from observed data. The aforementioned identification assumptions are required to express $\Psi$ - the difference in average unobserved potential outcomes - in terms of $E[Y \mid A=1, L] - E[Y \mid A=0, L]$ - the average difference in conditional outcome means between the two treatment groups. Identification refers precisely to this process of expressing (``identifying'') estimands such as $\Psi$ in terms of \textit{observed} data. In this setting with a single baseline treatment, the standard $IA$s\cite{Robins1986} are
\begin{itemize}
    \item[]$IA.1$: Conditional ignorability: $Y^a \perp A \mid L $.
    \item[]$IA.2$: Consistency: $P(Y^a = Y \mid A=a)=1, \ \forall a$.
    \item[]$IA.3$: No interference: $Y_i^{a_{1:n}} = Y_i^{a_i}$.
    \item[]$IA.4$: Positivity: $0<P(A=1 \mid L) < 1, \ \forall L\in\mathcal{L}$.
\end{itemize}
Above, $a_{1:n} = (a_1, a_2,\dots, a_i , \dots, a_n)$ is a vector of treatment interventions for each of $n$ observed subjects and $Y_i^{a_{1:n}}$ represents subject $i$'s potential outcome had each subject received their corresponding treatments in $a_{1:n}$. $IA.1$ requires that pre-treatment variables $L$ fully capture the confounding between treatment and outcome. That is, conditioning on $L$ renders the potential outcome under a particular treatment, $Y^a$, independent of the observed treatment assignment, $A$. This can be violated if, for instance, we fail to condition on some confounder, such as age, when in fact older subjects tend to be treated with treatment $A=1$ and are likely to have worse outcomes under this treatment, $Y^1$. It is important to note that conditioning on inappropriate variables (e.g. colliders or post-treatment variables) may also lead to ignorability violations\cite{Greenland1999}. In this paper, we will discuss ways to perform sensitivity analyses around violations of this assumption.

Consistency, $IA.2$, requires that the treatment be well-defined in terms of a clear intervention \cite{cole:fran:2009}. For example, suppose $A$ is high/low blood pressure and $Y$ is myocardial infarction. The outcome that occurs in a world where we intervene directly to lower blood pressure is likely not the outcome that would have occurred had everyone in the population had low blood pressure. This is because the mechanism by which we set blood pressure likely itself affects the outcome. Whether blood pressured was lowered via changes in lifestyle (exercise, better eating habits, etc) versus medication probably impacts the outcome. For this reason, consistency is often described as requiring that there is only ``one version'' of the treatment. A more well-defined intervention may be blood pressure medication use (versus no use).  Other canonical examples of ill-defined exposures include race and BMI \cite{hernan2008}.

$IA.3$ states that no subject's treatment assignment should affect another's potential outcome. Formally, the $i^{th}$ subject's potential outcome under intervention $A_i=a$, $Y_i^{a_i}$, need not be indexed by the other $n-1$ subjects' interventions in the superscript. Hence, we can simplify $Y_i^{a_{1:n}}$ to just $Y_i^{a_i}$. This assumption is often violated if subjects are not independent. For example, a study of the effect of prophylactic antivirals on infection using data from patients in the same hospital may suffer from interference: the antiviral treatment of subjects roomed together affect each other's infection probability. For concreteness, consider two such patients, $i$ and $j$, with potential infection status, $Y_i^{a_i, a_j}$ and $Y_j^{a_i, a_j}$, respectively. Here, intervention $a=1$ indicates antiviral therapy and $a=0$ indicates control. If the infection is contagious, it may be the case that $P(Y_i^{0, 1}) < P(Y_i^{0, 0})$. Even if subject $i$ is untreated, their infection probability would likely be lower had their neighbor, subject $j$, been treated. Since we cannot speak of subject $i$'s outcome separately from subject $j$'s treatment, we cannot drop $a_j$ from the superscript in $Y_i^{a_i, a_j}$. Causal inference in these settings is more complicated and an active area of research \cite{Hudgens2008}.

Finally, $IA.4$ requires that the treatment probability be bounded so that there is no subset of the population in terms of $L$ for whom treatment is deterministic. Intuitively, if treatment assignment was deterministic for a subpopulation of individuals, we would have no data about that group's outcome under the other treatment condition. Positivity violations can occur at the population level (e.g. protocols forbidding treatment $a$ for subjects over a certain age) or at the sample level due to small sample size (e.g. we observe no male diabetics with treatment $a$). The former are sometimes called \textit{structural} violations and the latter are called \textit{random} violations of positivity in the literature.

Using these assumptions we can identify both expectations in $\Psi$. First, under $IA.3$, $E[Y_i^{a_{1:n}}] = E[Y_i^{a_i}]$. Omitting subscripts for compactness, 
\begin{equation} \label{eq:std}
    \begin{split}
        E[Y^{a}] & = \int_{\mathcal{L}} E[Y^{a} \mid L] \ dP(L)  \\
                 & = \int_{\mathcal{L}} E[Y^{a} \mid A=a, L] \ dP(L) \\
                 & = \int_{\mathcal{L}} E[Y \mid A=a, L] \ dP(L) \\
    \end{split}
\end{equation}
The first equality follows from iterated expectation over $L$. We use $\mathcal{L}$ to denote the space of $L$. From $IA.1$, we know that the potential outcome is independent of assignment conditional on $L$, which allows us to condition on $A=a$ in the second equality. $IA.4$ ensures that we are not conditioning on a zero-probability event. Lastly, $IA.2$ allows us to drop the superscript. Intuitively, \eqref{eq:std} identifies the average potential outcome as a regression model (under intervention $A=a$) that is averaged over the marginal confounder distribution. Computing marginal causal effect using this expression is called \textit{standardization}. In this way, we have identified each term of $\Psi$ in terms a regression that is estimable from observed data.

\subsection{Statistical Assumptions}
Equation \eqref{eq:std} usually requires statistical/modeling assumptions about the regression, $E[Y \mid A, L]$. As an example, consider substituting a linear regression model $ E[Y\mid A, L] = \theta A + L'\beta $ (where an intercept is included in $L$). Then, under the $IA$s, standardization yields

\begin{equation}
	\label{eq:linearcase}
   \begin{split}
    \Psi & = \int_{\mathcal{L}} E[Y \mid A=1, L, \omega] - E[Y \mid A=0, L, \omega]dP(L) \\
         & = \int_{\mathcal{L}}\{ (\theta + L_i'\beta) - (L_i'\beta)\}dP(L) = \theta.
    \end{split}
\end{equation}
In this special case, the ATE, $\Psi$, is equal to the treatment indicator coefficient, $\theta$. Thus, an estimate of this coefficient is an estimate of the ATE. In the non-linear examples discussed later, $L_i'\beta$ will not cancel out as it did above and a probability model for $p(L)$ will be necessary to evaluate the integral.

\subsection{Bayesian Modeling}
Bayesian causal inference combines Bayesian modeling with the $IA$s discussed to compute a posterior distribution over causal estimands. In this section we introduce these key ideas, which will be expanded in future sections. Throughout much of the paper, we assume that $IA$s hold to keep focus on the added benefit of Bayesian modeling.

Suppose we observe data $D=\{Y_i, A_i, L_i \}_{i=1:n}$ on $n$ independent subjects, where $A_i\in\{0,1\}$ is a binary treatment indicator, $L_i$ is a vector of confounders (including an intercept), and $Y_i$ is the scalar outcome of interest, as defined earlier. Bayesian inference requires both a probability model for the conditional distribution of the outcome, $Y$, (a likelihood) as well as a probability distribution over the unknown parameter vector, $\omega$, governing this conditional distribution (i.e. a prior). Inference then follows from making probability statements about $\omega$ having conditioned on $D$ (via the posterior). From Bayes' rule we have that the posterior is proportional to the likelihood times the prior, $p(\omega \mid D) \propto p(\omega)\prod_i p( Y_i \mid A_i, L_i, \omega)$. 

For instance, if $Y_i$ is real-valued, we could specify a Gaussian outcome model with conditional mean $ \theta A_i + L_i'\beta $ and variance $\phi$: $p(Y_i \mid A_i, L_i, \omega) = N(Y_i \mid \theta A_i + L_i'\beta, \phi)$. We could also set a Normal-Inverse-Gamma prior on the parameter vector $\omega = (\theta, \beta, \phi)$, e.g. $p(\omega) = N(\theta \mid 0, 1)N(\beta \mid \mu_0, \Sigma_0) IG(\phi \mid a_0, b_0)$. This probability model induces a linear regression $E[Y\mid A_i, L_i] = \theta A_i + L_i'\beta $, where we drop explicit conditioning on the parameters. Now it remains to find the posterior over the model parameters, $p(\omega \mid D)$ - which includes $\theta$. As we showed with the linear model in the previous section, the coefficient  $\theta$ is the ATE, $\Psi$. So a posterior over $\theta$ is a posterior over $\Psi$. This simple example demonstrates a general Bayesian approach to causal inference. First, identify the causal parameter of interest as a transformation of the model parameters. The $IA$s required to achieve this will vary by problem and strategy. Mediation \cite{imai2010} and time-varying treatment  \cite{Daniel2013} settings will require extensions of the $IA$s discussed. Instrumental variables \cite{Baiocchi2014}, difference-in-differences \cite{Lechner2010}, and regression discontinuity \cite{Imbens2007} strategies all involve their own unique $IA$s. Second, obtain the posterior distribution (or draws from it) of these model parameters which, after transformation, yields a posterior over the causal estimand.

In practical settings, the posterior distribution, $p(\omega \mid D)$ does not have known form - so that we cannot analytically find the posterior after specifying the likelihood and prior. As a result, inference is instead typically conducted using \textit{draws} from the posterior obtained via Markov Chain Monte Carlo (MCMC). Though a crucial topic and active area of research in itself, we omit discussion of MCMC methods and keep focus on Bayesian estimation of causal effects. We refer the reader to Andrieu et al. \cite{andrieu2003} for an introduction to MCMC. For our purposes, it is enough to know that MCMC yields a set of $M$ draws, $\{\omega^{(m)}\}_{1:M}$, from the posterior $p(\omega\mid D)$ given a specified likelihood and prior. Throughout, we assume we have sufficiently many draws to closely approximate the posterior.  The mean or median of these samples can be used as a Bayesian point estimate of $\omega$. Percentiles of these draws can be used for credible interval estimation (e.g. .025 and .975 percentiles for a 95\% interval). 

This paper relies mainly on \textbf{Stan} for MCMC sampling throughout. \textbf{Stan} is an open-source programming language for specifying Bayesian models using intuitive syntax. It back-ends to C++ to efficiently obtain posterior draws after a likelihood and prior are specified. \textbf{Stan} programs are often called in \textbf{R} via the package \textbf{rstan}. For those unfamiliar with \textbf{Stan} and \textbf{R}, we provide some guidance with \textbf{SAS} version 9.4 - a popular commercial statistical analysis software. Some of the nonparametric models to be discussed cannot be handled in either \textbf{Stan} or \textbf{SAS}. For these models, we will rely on specialized \textbf{R} packages.

\subsection{Prior Information}

As mentioned earlier, Bayesian inference requires specification of a prior over the parameters, $p(\omega)$. Throughout this paper we hope to illustrate that, rather than anchoring estimates to particular hard-coded values, priors can induce intricate correlation structures between parameters. These correlation structures stabilize causal effect estimates when data are sparse (as it often is in scientific applications). This is often referred to as ``shrinkage''. Priors can also be used to induce ``sparsity'' on whole parameter vectors. Specifically, for high-dimensional vectors, we can place priors that express the belief that some portion of the vectors are nearly zero. Priors can also be used to conduct probabilistic sensitivity analyses around causal identification assumptions. All of these are pragmatic motivations for taking a Bayesian approach to causal estimation, even if one is not a Bayesian ``at heart''. We will emphasize that, perhaps contrary to intuition, common frequentist approaches can often be seen as special cases of these Bayesian estimators with very rigid priors.

\section{Parametric Models in Point-Treatment Settings}
In the following sections, we outline two examples where a Bayesian approach to causal inference offers unique benefits in the form of prior shrinkage. Although these examples use relatively simple parametric models, they reflect the general approach and intuition of Bayesian causal inference and help motivate key tools such as the Bayesian bootstrap.

\begin{figure*}[t]
    \centering
    \begin{subfigure}[b]{.49\textwidth}
        \centering
        \includegraphics[width=\textwidth]{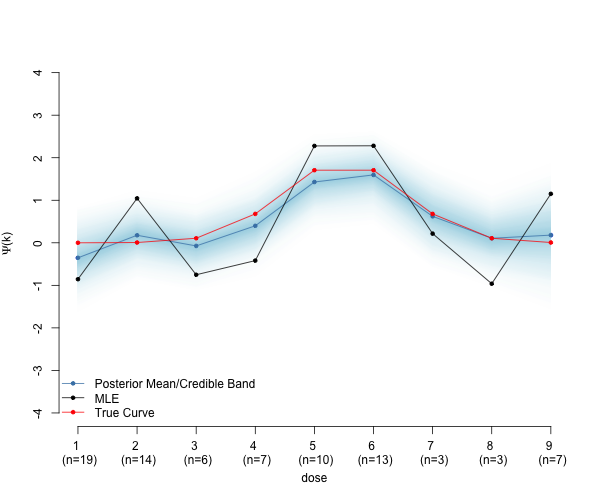} 
        \caption{Posterior estimates from \eqref{eq:drmodel} with prior \eqref{eq:ar1prior} with $K=10$, $\mu_1=0$, $\tau_1=10$, and $\tau_k=1$ for all $k$. The AR1 prior smooths erratic MLEs by inducing correlation between neighboring points on the curve.}
        \label{fig:doseresponse}
    \end{subfigure}
    \hfill
    \begin{subfigure}[b]{.49\textwidth}
        \centering
        \includegraphics[width=\textwidth]{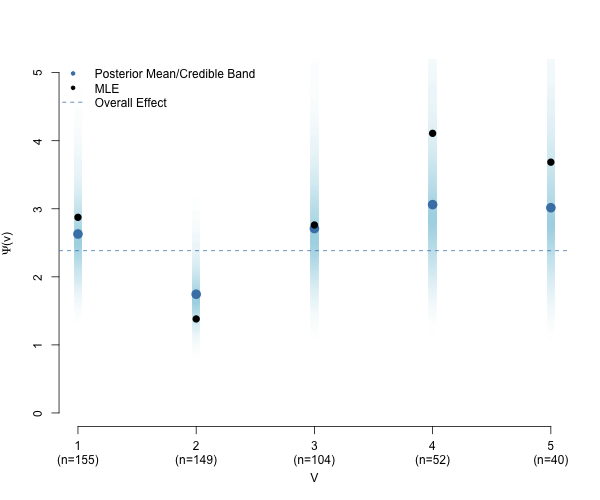} 
        \caption{Posterior estimates of $\Psi(v)$ from \eqref{eq:logit} with partial pooling prior of Section \ref{sc:logistic} with $q=5$, a single confounder $W$, an improper uniform prior on $\mu$, and $\tau=.5$. Posterior mean odds ratio for each stratum are shrunk towards the overall causal odds ratio (dotted line).} 
        \label{fig:partialpooling}
    \end{subfigure}
    \vspace{.1in}
    \caption{Estimates of dose effect curve (Section \ref{sc:doseresponse}) and partially pooled conditional causal odds ratios (Section \ref{sc:logistic}) using synthetic data.}
\end{figure*}

\subsection{Causal Dose Effects with AR1 Prior} \label{sc:doseresponse}

Consider a setting where treatment consists of $K$ dose levels $A_i \in \{0,1, \dots, K \}$, with $A_i=0$ indicating no treatment. Let $A_{ik} = \I(A_i=k)$ be an indicator that subject $i$ was assigned to dose $k\in\{1,\dots, K\}$. Here, we assume the dose values are ordered so that they are increasing with $k$. That is, dose $k+1$ is higher than dose $k$. Consider a linear outcome model,
\begin{equation} \label{eq:drmodel}
    \begin{split}
         E[Y_i \mid A, L_i] = \theta_0 + L_i' \beta  + \sum_{k=1}^{K} \theta_k A_{ik}. 
    \end{split}
\end{equation}
Suppose our estimand of interest is a causal incremental dose effect curve on $\Psi(k) = E[Y^{A=k}] - E[Y^{A=k-1}] $. This is a curve as a function of dose, $k$. Each point on the curve is the causal effect of increasing dose from level $k-1$ to level $k$. Under mild extensions of $IA.1$-$IA.4$ from the binary treatment setting to the multi-treatment setting we can again identify this estimand as 
$$ \Psi(k) = \theta_k - \theta_{k-1} \ \ k\in\{2, \dots, K \} $$
Where the first point is $\Psi(1)=\theta_1$. We consider several prior choices for $\theta_{1:K}$ and the induced prior on $\Psi(k)$. Throughout, $u_{a:b}$ for intergers $a<b$ denotes the vector $u = (u_a, \dots, u_b)$.  A first-pass approach may be to express prior independence and factorize the joint prior as $p(\theta_{1:K}) = \prod_{k=1}^K p(\theta_k)$. We could specify each term to be Gaussian centered at some prior mean, $\mu_k$, and standard deviation, $\tau_k$. However, we can formulate more useful priors in this setting. The increasing dose levels may give us prior reason to believe that the effect of neighboring doses are actually correlated, not fully independent. This motivates an alternative (dependent) prior factorization: $p(\theta_{1:K}) = p(\theta_1)p(\theta_2\mid \theta_1) \prod_{k=3}^K p(\theta_k \mid \theta_{K-1}, \theta_{K-2})$. Each term is specified as 
\begin{equation*}
    \begin{split}
        \theta_1 \sim &  N(\mu_1, \tau_1) \\ 
        \theta_2\mid \theta_1 \sim &  N(2\theta_1, \tau_2) \\
        \theta_k \mid \theta_{k-1}, \theta_{k-2} \sim & N(2\theta_{k-1} - \theta_{k-2}, \tau_k), \ \ \  k > 2,
    \end{split}
\end{equation*}
where $\mu_1$, $\tau_{1:K}$ are all hyperparameters that we can specify. Alternatively, we could specify hyperpriors for these parameters. The above induces the following first-order autoregressive (AR1) prior on the causal curve, $\Psi(k)$. For instance, the last line for $k>2$ above implies that $\theta_k - \theta_{k-1}  \mid \theta_{k-1}, \theta_{k-2} \sim  N(\theta_{k-1} - \theta_{k-2}, \tau_k)$. This follows from simply subtracting $\theta_{k-1}$ from $\theta_k$ and its mean. Using the definition of $\Psi(k)$, we see that this statement is equivalent to $\Psi(k)  \mid \theta_{k-1}, \theta_{k-2} \sim N( \Psi(k-1), \tau_k)$. Extending this logic, the hierarchical prior on $\theta$s induces the following prior on the $\Psi$s
\begin{equation} \label{eq:ar1prior}
    \begin{split}
        \Psi(1) & \sim N(\mu_1, \tau_1) \\
    \Psi(k) \mid \Psi(k-1) & \sim N(\Psi(k-1), \tau_k), \ \ k>1
    \end{split}
\end{equation}
This expresses the prior belief that the response from increasing dose to the next level should not be too different from the response due to the previous dose level. That is, neighboring points on the curve are related. Of course, if we have data suggesting otherwise, the data will drive our posterior inference away from this prior. However, in the absence of data, this provides valuable shrinkage back towards a sensible prior belief. An example using synthetic data is presented in Figure \ref{fig:doseresponse} with posterior sampling done in \textbf{Stan} \cite{Stan}. Implementation details along with a more thorough walkthrough using this synthetic data set are available in Appendix \ref{ap:dose_response}. Implementation via PROC MCMC in \textbf{SAS} is also discussed. Notice in the figure that small sample sizes at each dose level lead to erratic MLE estimates. In contrast, the Bayesian estimate with the AR1 prior produces a smoother curve. In dose level 8, we only have three observations. Thus, the Bayes estimate is aggresively shrunk towards the estimate at dose 7.

A common heuristic solution to this issue of decreasing sample size with increasing dose is to fully pool patients at, say, dose $K$ and $K-1$ and estimate a single effect for both rather than allowing separate effects. The prior in \eqref{eq:ar1prior} is a compromise between these two extremes. Recall from \eqref{eq:ar1prior} that $\Psi(K) \mid \Psi(K-1) \sim N(\Psi(K-1), \tau_K)$ for $K>1$. Now notice that the heuristic alternative of combining groups $K$ and $K-1$ corresponds to the strong prior belief that $\tau_K \approx 0$. That is, the causal effect at dose $K$ is a point-mass distribution at $\Psi(K-1)$:  $\Psi(K) \mid \Psi(K-1) \sim \delta_{\Psi(K-1)}$.

\subsection{Partial Pooling of Conditional Causal Effects} \label{sc:logistic}

Here we consider a more involved model for causal estimation using a logistic regression with binary outcome and treatment. Here, the mean function $E[Y \mid A, L]$ is related to the covariates $L$ through a (non-linear) logit link with inverse logit denoted by $\sigma\{\cdot\}$. Thus, the integration over $L$ in \eqref{eq:std} must be evaluated explicitly. Consider some $q-$dimensional subset of pre-treatment covariates, $V \subset L$. Let $W= L\setminus V$ be the set difference so that $L=\{W, V \}$. One target estimand of interest in this setting is a causal odds ratio at each level of $V$ 
\begin{equation}\label{eq:causalOR}
    \begin{split}
    \Psi(v) & = \frac{ E[Y^1\mid V=v] / (1 - E[Y^1 \mid V=v])}{E[Y^0\mid V=v] / (1-E[Y^0\mid V=v])}
    \end{split}
\end{equation}
Causally, this is contrasting the odds of the outcome had everyone in stratum $V=v$ taken treatment $1$ versus treatment $0$. Since we allow for different treatment effects for each level of $v$, sometimes the set of $\Psi(v)$ is referred to as a \textit{heterogenous treatment effects} or  \textit{conditional treatment effects} (i.e. conditional on $V=v$). Under extensions of $IA.1-IA.4$, each conditional expectation in $\Psi(v)$ is identified as $E[Y^a \mid V=v] =\int_{\mathcal{W}}E[Y \mid a, V=v,W]dP_v(W)$. Where $P_v(W) = P(W \mid V=v)$ is the confounder distribution within stratum $v$. Note that this is just \eqref{eq:std} conditional on $V=v$. The regression model is  
\begin{equation} \label{eq:logit}
    E[Y \mid A=a, V, W] = \sigma\big\{\beta_w'W + \beta_v'V + (\theta_0 + \theta_{1:q}' V) A  \big\}
\end{equation}

Above, we include an intercept in $W$. Note that the treatment effect, $\theta_0 + \theta_{1:q}' V$, varies with levels of $V$. We defer discussion of the integration over $W$ to Section \ref{sc:bb}. For concreteness, suppose $V$ is a vector of indicators for $q=4$ race/ethnicity categories: Black, Asian, Hispanic, Native American, and White as reference. Often some categories (e.g. Hispanic, Asian, Native American) are sparse. In these settings, it is common to combine these categories into ``Other'' and estimate a single odds ratio for these subjects. It is also common to simply exclude these subjects and not compute causal effects for these strata at all. Neither of these may be desirable and, again, carefully formulated priors can help us strike a balance when estimating conditional causal effects. For instance, consider the prior assumption that all of these \textit{conditional} (we have not marginalized over $W$ yet) effects within race category ($\theta_0$, $\theta_0 + \theta_1$, \dots, $\theta_0 + \theta_4$) are normally distributed around some ``overal'' treatment effect $\mu$ with standard deviation $\tau$. We can achieve this by specifying a Gaussian prior for the conditional effect in the reference stratum $\theta_0 \sim N(\mu, \tau)$. For the conditional effect in stratum $V=1$, we specify $\theta_0 + \theta_1 \sim N(\mu, \tau)$. This is the same as saying $\theta_1 \sim N(\mu - \theta_0, \tau)$.  Following this logic for the other strata, the joint prior over all parameters is 
\begin{equation}
    \label{eq:partialpoolprior}
    p(\theta_{0:4}\mid \mu, \tau) = N(\theta_0\mid \mu, \tau) \prod_{j=1}^4 N(\theta_j \mid \mu - \theta_0, \tau)
\end{equation}
For categories with many observations, the posterior of the conditional effects with race category will be driven mostly by data. However, for small categories, each conditional effect shrinks to the overall average across race values, $\mu$. The hyperparameter $\tau$ controls how aggressively we shrink these conditional effects to the overall average. This allows us to estimate \textit{regularized} race-specific causal effects rather than abandoning the task altogether or resorting to ad-hoc groupings of categories. Priors for $\mu$, $\tau$, and the other regression coefficients must be specified. Standard guidance \cite{gelmanbda04} can be followed when specifying priors on these nuisance parameters. Similar to the dose effect example, the heuristic approach of fully pooling sparsely populated race clusters corresponds to a rigid prior. In this case, a prior belief that the conditional effect in all the pooled strata are the same.

\subsection{Standardization via Bayesian Bootstrap} \label{sc:bb}

To compute conditional causal effects in  \eqref{eq:causalOR}, we must integrate the logistic regression in the previous example over $P_v(W)$. As shown earlier, factors involving $W$ would cancel out in a linear model - obviating the need for explicit integration. Here, due to the non-collapsability of the logit link, $W$ does not cancel. To compute this integral, we need an estimate of $P_v(W)$ over which to integrate. Let $S_v = \{ i : V_i =v \}$ be the set of indices of subjects in stratum $V=v$. There are $n_v$ subjects in stratum $v$, so that the size of $S_v$ is $n_v$.  A frequentist nonparametric approach would be to estimate the distribution empirically as $\hat P_v(w) = \frac{1}{n_v} \sum_{j \in S_v} \delta_{W_j}(w)$, where $\delta_{W_j}(\cdot)$ is the degenerate distribution at $W_j$. This places probability mass $1/n_v$ on each of the $n_v$ subjects in stratum $v$. This yields average potential outcome estimate
$$E[Y^a \mid V=v] \approx \frac{1}{n_v}\sum_{ j \in S_v } E[Y \mid A=a, V=v, W=W_j]$$
in stratum $v$. This is ideal in the sense that we impose no parametric model on the conditional distribution of $W$, but is unsatisfactory from a Bayesian point of view because it ignores the uncertainty in the empirical estimate. This motivates the Bayesian bootstrap (BB) \cite{Rubin1981}. The BB begins with a model for  $W$, $P_v(w ) =\sum_{j \in S_v } p_{vj} \cdot \delta_{W_j}(w) $. We store all the weights in this stratum in an $n_v-$dimensional vector $p_v = \{ p_{vj} : j \in S_v\}$. This weight vector $p_v$ live in a simplex $p_{v} \in \{ \mathbb{R}^{n_v} : p_{vj} >0 \  \forall j \text{ and } \sum_{j \in S_v} p_{v j}  = 1 \}$. Rather than fixing $p_{vj} = 1/n_v$, the BB treats the weights as unknown with a flat Dirichlet prior $p_{v} \sim Dir(0_{n_v})$, where  $0_{n_v}$ is the $n_v-$dimensional vector of zeros. This yields the (conjugate) posterior $p_{v} \mid W \sim Dir( 1_{n_v})$, where $1_{n_v}$ is the $n_v-$dimensional vector of ones. The BB makes minimal assumptions about the confounder distribution within each stratum: note the posterior mean of each weight is also $1/n_v$ (same as the frequentist approach), but allows uncertainty around this mean to flow through to the causal effect in that stratum. 

The BB was applied to marginal ATE estimation using generalized linear models (GLMs) for the outcome by Wang et al. \cite{Wang2015}.  When computing a marginal ATE, the BB is used as a model for the \textit{marginal} $p(L)$ not the conditional $p_v(W)$. In the outcome model, we no longer set $V=v$ since $V$ is included in $L=\{W,V\}$, which we integrate over. The BB model is now $P(l \mid p_{1:n}) =\sum_{i=1}^n  p_i \delta_{L_i}(l)$. Now we place a Dirichlet prior on the $n$-dimensional vector $p_{1:n}$ rather than the $n_v$-dimensional vectors $p_v$. This marginal estimate will play a key role in the nonparametric estimation of Section \ref{sc:bnp}.

Full posterior inference for the causal odds ratio \eqref{eq:causalOR} requires just a few additional steps after sampling. Suppose we obtain the $m^{th}$ draw of the parameters in \eqref{eq:logit}, $\{\beta_w^{(m)},\beta_v^{(m)}, \theta^{(m)}, \theta_{1:q}^{(m)} \}$. Then, for each stratum $v$, we draw BB weights $p_{v}^{(m)} \mid W \sim Dir(1_{n_v})$. Note that here $p_v^{(m)}$ denotes the collection $\{ p_{vj}^{(m)} : j \in S_v \} $. Then, we do the following
\begin{enumerate}
    \item \textbf{Integrate} under both interventions $A \in \{1,0\}$:
    \begin{equation*}
        \begin{split}
            \mu^{(m)}(a,v) & = \sum_{j \in S_v } p^{(m)}_{vj} \sigma\big\{\beta_w^{'(m)} W_j + \beta_v^{'(m)}v + (\theta^{(m)} + \theta_{1:q}^{'(m)} v)a  \big\} 
        \end{split}
    \end{equation*}
    \item \textbf{Compute Causal Effects} for each $v$
    $$ \Psi^{(m)}(v) = \frac{\mu^{(m)}(1,v)/(1-\mu^{(m)}(1,v))}{\mu^{(m)}(0,v)/(1-\mu^{(m)}(0,v))} $$
\end{enumerate}

Doing this for $m=1, \dots, M$ posterior parameter draws yields $M$ draws from the posterior of the causal estimand: $ \{ \Psi^{(m)}(v) \}_{1:M}$ for each $v$. These draws can be used for posterior inference.  Figure \ref{fig:partialpooling} shows posterior estimates of $\Psi(v)$ with the partial pooling prior in \eqref{eq:partialpoolprior} using synthetic data. MCMC-based posterior inference was done using \textbf{Stan}. Notice that for strata $V\in\{4,5\}$, we have relatively few observations. In these strata, the maximum likelihood estimate (MLE) is much higher than the others due to small sample variability. Thus, the Bayesian prior aggressively shrinks the posterior mean estimate away from MLE towards the overall effect. From a causal perspective, we can view this as shrinking the heterogenous treatment effects towards an overall treatment effect. Details of this synthetic data generation and implementation in both \textbf{Stan} and \textbf{SAS} are given in Appendix \ref{ap:partial_pool}. The latter relies on PROC MCMC for posterior sampling and PROC IML for the BB post-processing step.

\section{Time-varying Treatment and Confounding} \label{sc:gcomp}
The previous sections focused on the point-treatment setting: estimating the causal effect of a single treatment administered at baseline while adjusting for a single set of pre-treatment parameters. In many applications, treatment decisions are made sequentially over time as a function of covariates measured after baseline. For example, consider a binary treatment setting where treatment at time $t=0$, $A_0$, is assigned conditional on confounders, $L_0$, measured before $A_0$. The subsequent treatment, $A_{1}$, is assigned conditional on $L_0$, $A_0$, and $L_1$, where $L_1$ is measured between $A_0$ and $A_1$, temporally. After treatment, we observe a single outcome $Y$. Suppose we wish to estimate the causal ATE $ E[Y^{(1,1)} - Y^{(0,0)}]$ - the difference had everyone in the target population been always treated versus never treated. Note the potential outcomes here are indexed by a treatment \textit{vector}, not scalar. In the literature this vector is often referred to as a ``treatment regime'' or ``treatment policy''. In this section, we first discuss causal contrasts comparing outcomes that would have been realized under two different \textit{static} treatment regimes while controlling for time-varying confounding. Static regimes are treatment vectors that are pre-set to fixed values in advance (e.g. always treated, never treated, alternating treatments). Afterwards, we discuss an extension to \textit{dynamic} treatment regimes, where the treatment regime is set sequentially over time according to a pre-specified rule (e.g. treat at time point $t$ if blood pressure at time $t$ is lower than some threshold). We refer the reader to Daniel \cite{Daniel2013} for a thorough tutorial on time-dependent confounding and modeling.

\subsection{Comparing Static Treatment Regimes}
\label{sc:staticregime}

Standard regression methods fail to properly adjust for the time-varying confounder in these settings. For instance,  if we condition on $L_1$, then we adjust away $A_0$'s impact on $Y$ that runs through $L_1$. However, $L_1$ is a confounder of $A_1$ and $Y$ - so failing to adjust for it will also lead to bias. Now, generalizing to $t=0,\dots, T$ time points, under extensions of $IA.1-IA.4$ to this sequential setting \cite{Robins1986} we can identify each term of the causal contrast $\Psi = E[ Y^{a_{0:T}} - Y^{a_{0:T}'}]$  as
\begin{equation} \label{eq:gcomp}
        E[Y^{a_{0:T}}] = \int_\mathcal{L} E[ Y \mid a_{0:T}, L_{0:T}] \times \prod_{t=0}^{T} p(L_t \mid L_{0:t-1}, a_{0:t-1}) d L_{0:T}
\end{equation}
Where for $t=0$, we define $p(L_0 \mid L_{0:-1}, a_{0:-1})=p(L_0)$. The expression above is known as the $g$-formula and the computation of the integral is referred to as $g$-computation - it is the multi-time point generalization of standardization in \eqref{eq:std}. Here we ignore the details of identification to focus on Bayesian modeling and computation. 

\begin{figure*}
    \centering
    \begin{subfigure}{.49\textwidth}
        \centering
        \includegraphics[width=\textwidth]{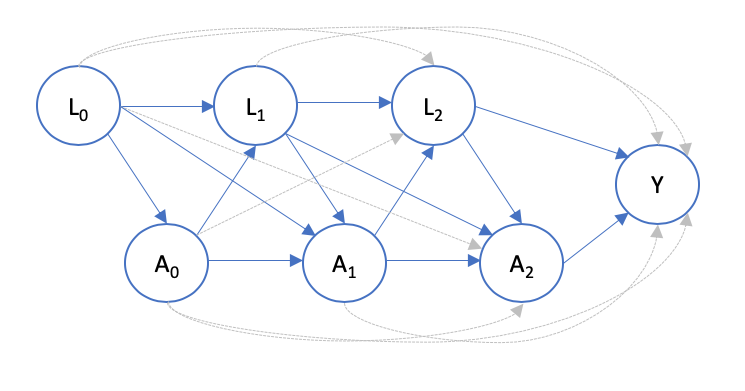} 
        \caption{ } 
        \label{fig:dagfull}
    \end{subfigure} \hfill
    \begin{subfigure}{.49\textwidth}
        \centering
        \includegraphics[width=\textwidth]{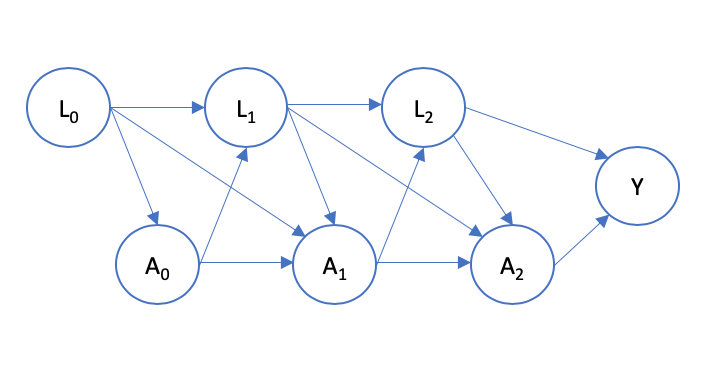} 
       \caption{ } 
        \label{fig:dagmarkov}
    \end{subfigure}
    \vspace{.1in}
    \caption{A directed acyclic graph (DAG) showing a time-varying treatment, $A_t$, time-varying confounder $L_t$, and outcome, $Y$, for three time points. In the first panel, treatment and confounding at each time point affects treatment and confounding in \textit{every} future time point. The second panel depicts the Markov assumption described in Section \ref{sc:staticregime} - confounders and treatment only impact variables in the next period so that $p(L_2 \mid L_{0:1}, A_{0:1}) = p(L_2 \mid L_{1}, A_{1})$. This is visually depicted by the deletion of the gray arrows in the first panel. Bayesian methods can help us strike a balance between these two extremes. }
\end{figure*}

In particular, note that the above requires integrating an outcome regression over the joint distribution of confounders, conditional on treatment regime $a_{0:T}$. The outcome regression here can be high-dimensional even in common data applications. If we have just two time-varying confounders and twelve (e.g. monthly) time points, the outcome model must condition on 36 variables. Similarly, each conditional confounder distribution must (usually) be modeled conditional on \textit{all} previous values of $L_t$ and $A_t$ - another high-dimensional modeling task.  

The sequential nature of treatment and confounder measurement can be visually depicted using a directed acyclic graph (DAG) in Figure \ref{fig:dagfull} for $T=3$ timepoints (for compactness). Notice $L_2$ is impacted by \textit{all} previous confounder values ($L_0$ and $L_1$) and treatment values ($A_0$ and $A_1$). This is shown by arrows going into $L_2$. Similarly, the outcome is impacted by \textit{all} past $L$ and $A$ values. To simplify this complexity, a Markovian assumption is commonly invoked. This assumes that each confounder distribution only depends on the previous confounder and treatment values, $p(L_t \mid L_{0:t-1}, A_{0:t-1}) = p(L_t \mid L_{t-1}, A_{t-1})$. A similar assumption may be used in the outcome model. This Markov-type assumption is depicted in Figure \ref{fig:dagmarkov}, which is simply the DAG in \ref{fig:dagfull} with all the gray arrows removed. After removing gray arrows, each variable is \textit{directly} impacted only by variables in the preceding time point. In Figure \ref{fig:dagmarkov}, for instance, once we know $L_1$ and $A_1$, we know the distribution of $L_2$. The history ($A_0$ and $L_0$) need not be considered since it only affects $L_2$ through $L_1$ and $A_1$. Thus, $p(L_2 \mid L_{0:1}, A_{0:1}) = p(L_2 \mid L_{1}, A_{1})$. 

Neither of these extremes - conditioning on full history or invoking Markov - are completely desirable. Suppose $L_t$ is an indicator of poor kidney function at day $t$. The Markov assumption presumes that two treated subjects with, say, poor kidney function on the previous day, $L_{t-1}$, have the same $L_t$ distribution - even if one patient had poor function everyday since $t=0$ and the other had good function until day $t-1$. This seems unrealistic. On the other hand, it may also be unrealistic to say that kidney function on day 1, $L_1$, would directly impact function on, say, day 100, $L_{100}$. In the Bayesian paradigm, we can devise priors that balance these two extremes of either conditioning only on previous time period's values versus conditioning on the entire past history. The general idea is to condition on the full history, but express a prior belief that values closer in time to the present have relatively more direct impact on the present. Conversely, values further back in time have small, if any, direct effect.

\begin{figure*}
    \centering
    \begin{subfigure}{.49\textwidth}
        \centering
        \includegraphics[width=\textwidth]{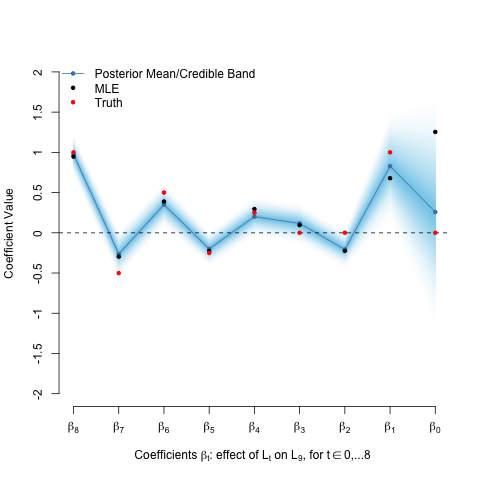} 
        \caption{ Plot of coefficient estimates from \eqref{eq:condconf} with t=9. Each coefficient on the x-axis is the effect of $L_t$ on $L_9$ for time points $t=0,\dots, 8$. Note aggressive shrinkage of $\beta_0$ but ability to detect signal in the past at $\beta_1$.} 
        \label{fig:gcomp_sparse}
    \end{subfigure} \hfill
    \begin{subfigure}{.49\textwidth}
        \centering
        \includegraphics[width=\textwidth]{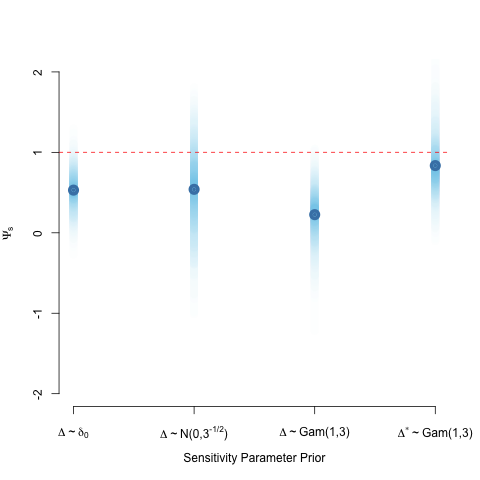} 
        \caption{ Sensitivity analysis of Section \ref{sc:sensitivity}. Posterior Distribution of $\Psi_s = \Psi - \Delta$ under various priors for for $\Delta$. Red line indicates true value. } 
        \label{fig:sensitivity}
    \end{subfigure}
    \vspace{.1in}
    \caption{Example of $g$-computation on synthetic data with 10 time points, single time-varying treatment and confounder. The ridge prior in \eqref{eq:ridge} was used along with Gaussian outcome and conditional confounder models.}
\end{figure*}

To continue this example, consider a simple setting where $L_t$ is a continuous measure of kidney function and the outcome is viral load, with the treatment, $A_t\in \{0,1\}$ being anti-viral therapy at time $t$. Lower viral load is desirable, but comes at the expense of nephrotoxicity. So, depending on previous treatment, if the patient shows poor kidney function as measured by $L$, the physician may alter their treatment. To evaluate \eqref{eq:gcomp}, we will need an outcome regression and a sequence of conditional confounder models. Consider a linear outcome regression $E[Y \mid A_{0:T}, L_0:T] = \gamma + L_{0:T}'\beta_Y + A_{0:T}' \theta_Y$ and a Gaussian conditional model for $L_t$ with conditional mean
\begin{equation} \label{eq:condconf}
    \mu_{L_t}( L_{0:t-1}, A_{0:t-1}) = \beta + L_{0:t-1}' \beta_L  + A_{0:t-1}' \theta,
\end{equation}
where $\beta_L = (\beta_{0}, \beta_{1}, \beta_2, \dots, \beta_{t-1})$ and $\theta$ are length $t$ parameter vectors. We note that these parameters should be indexed by $t$ (e.g. $\beta_L^t$) as each conditional distribution should be allowed to have their own effects, but we omit this indexing for compactness. We consider the following prior on each element of $\beta_L$
\begin{equation} \label{eq:ridge}
    \beta_{t-k} \sim N(0,  \tau_k \phi ), \ \ k\in\{1,\dots, t \}.
\end{equation}
An identical prior can be used for $\theta$. Consider the specification $\tau_k = (1/\lambda^k)$ for some $\lambda>1$. This corresponds to what is often referred to as a ridge penalty in the machine learning literature. However, it differs from the standard ridge regression in that we do not apply the same penalty to all coefficients. Rather, the penalty gets increasingly aggressive for coefficients going farther back in time. For instance, for $\lambda=2$, the prior standard deviation around 0 is halved every step backward in time, providing increasingly aggressive prior shrinkage towards 0. This implies a strong prior belief that recent confounder values are more likely to influence the present than values farther in the past. Note that the Markov assumption follows from a special (strongly informative) case of this prior, where $\beta_{t-k} \sim \delta_0$ for $k >1$: all coefficients but $\beta_{t-1}$ follow a point-mass distribution at 0. An example of the prior in \eqref{eq:ridge} is provided in Figure \ref{fig:gcomp_sparse} with $T=9$. The plot shows the coefficients of $\beta_L$ in the model $\mu_{L_9}$ getting increasingly penalized. Note that the posterior estimate of $\beta_1$ is able to break away from this prior to detect a signal (a truly non-zero coefficient value), even though it is farther in the past. However, at time point $0$ the posterior estimate $\beta_0$ is strongly shrunk to zero (relative to the MLE). 

The Bayesian literature has explored several such ``sparsity'' priors, including the horseshoe \cite{Carvalho2010}, LASSO, and spike-and-slab priors \cite{George1997} - all of which could be applied to $g$-computation. These priors can all be characterized by their ability to both shrink noise, while being able to break away from the prior to detect signals \cite{Carvalho2010}. For instance, a horseshoe type prior on the components of $\beta_L$ can be specified by placing half-Cauchy hyper-priors on $\tau_k$ and $\phi$ in \eqref{eq:ridge}: 
\begin{align*}
    \tau_k \sim &  C^+(0, 1/2^k) \\
    \phi \sim & C^+(0, \nu)
\end{align*} 
where $\nu$ is a specified scale parameter that controls overall shrinkage across time. Similar to the ridge-type prior, the scale on the distribution of $\tau_k$ is halved every step backward in time.

The integration in \eqref{eq:gcomp} can be done via Monte Carlo after obtaining MCMC draws from the posterior of all the parameters governing the conditional confounder and outcome distributions. Conditional on these draws, we can simulate  confounder values from these distributions and take the average of our regression model over these simulated values.  Let $\omega_y^{(m)}$ denote the $m^{th}$ draw of the parameter vector governing the regression in \eqref{eq:gcomp}. Similarly, denote the parameter vector governing each conditional confounder distribution $p(L_t \mid L_{0:t-1}, a_{0:t-1})$ by $\omega_{L_t}^{(m)}$. For instance, these would include draws of the regression parameters $\beta, \beta_L, \theta$ in \eqref{eq:condconf} along with the Gaussian variance parameter. In the viral load example discussed earlier, $\omega_y^{(m)}$ would consist of $\omega_y^{(m)} = (\gamma^{(m)}, \beta_Y^{(m)}, \theta_Y^{(m)}) $

To compute the causal ATE $\Psi = E[Y^{a_{0:T}} - Y^{a_{0:T}'}]$ of regime $a_{0:T}$ versus $a_{0:T}'$, with each posterior draw of $\omega_Y^{(m)}$ and $\omega_{L_t}^{(m)}$ we:
\begin{enumerate}
    \item \textbf{Draw confounders}, for $t \in 0\dots T$ sequentially
    $$ \tilde L_t \sim p(L_t \mid \tilde L_{0:t-1}, a_{0:t-1}, \omega_{L_t}^{(m)})$$
    Denote these draws $\tilde L_{0:T} = ( \tilde L_0, \tilde L_1 \dots,  \tilde L_T)$. Repeat this a total of $B$ times to obtain $ \{  \tilde L_{0:T}^{(b)}\}_{1:B} = \{ \tilde L_{0:T}^{(1)}, \tilde L_{0:T}^{(2)}, \dots \tilde L_{0:T}^{(B)} \}$
    \item \textbf{Integrate the outcome model} $E[Y \mid a_{0:T}, L_{0:T} ]$ over $ \{\tilde L_{0:T}^{(b)}\}_{1:B}$ conditional on current set of draws $\omega_y^{(m)}$, under both interventions. In the viral load example, this would be 
    \begin{equation*}
        \begin{split}
            \mu^{(m)}(a_{0:T}) & =  \frac{1}{B}\sum_{b=1}^B E[Y \mid a_{0:T}, \tilde L_{0:T}^{(b)}, \omega_y^{(m)}] \\
                               & = \frac{1}{B} \sum_{b=1}^B (\gamma^{(m)} + \tilde L_{0:T}^{'(b)}\beta_Y^{(m)} + (a_{0:T}) ' \theta_Y^{(m)})
        \end{split}
    \end{equation*}
    Similarly, repeat Step 1 and 2 under $a_{0:T}'$. 
    \item \textbf{Compute Causal Contrast}
    $$ \Psi^{(m)} =\mu^{(m)}(a_{0:T}) - \mu^{(m)}(a_{0:T}')$$
\end{enumerate}
This procedure yields $M$ posterior draws of $\Psi$, which can be used to form posterior mean and credible intervals. This can also be implemented in \textbf{Stan} using the ``generated quantities'' block as demonstrated in the our code on GitHub. The number of draws $B$ should be large so that the Monte Carlo error of the integration of \eqref{eq:gcomp} is sufficiently low. In general, analyses with more time points and time-varying confounders will require larger $B$. In practice, we can try running steps 1-3 for a single posterior draw (say, draw $m$), over successively larger $B$. Keeping track of each repetition, we can check at which point increasing $B$ only marginally increases precision in the estimate of $\Psi^{(m)}$. We can then set $B$ to this value across all posterior draws. A nice feature of this Bayesian approach is that uncertainty about the confounder and outcome models at all time points naturally flows through to the posterior of $\Psi$ or any other causal contrast. For instance, we could have computed posterior draws of causal ratio contrast $E[Y^{a_{0:T}}]/E[Y^{a_{0:T}'}]$ in Step 3 as $\mu^{(m)}(a_{0:T}) / \mu^{(m)}(a_{0:T}' )$. In contrast, the frequentist approach would require many bootstrap estimates of the parameter vectors. Then, we would repeat Steps 1-3 using these bootstrap draws in place of the posterior draws. In the Bayesian framework, we need not re-estimate the model. We simply post-process the same set of draws differently.

\subsection{Dynamic Treatment Regimes}
In the previous section we compared static treatment vectors $a_{0:T}=(a_0, a_1, \dots, a_T)$, where each element is fixed at baseline. A dynamic treatment regime is a treatment regime where the elements are determined dynamically post-baseline via a pre-specified decision rule. A decision rule is a function that, at each time point $t$, maps the confounder history and treatment history $(L_{0:T}, A_{0:T})$ to a treatment value $a_t \in \{0,1\}$. For simplicity, here we discuss treatment rules that determine assignment based on current confounder values only. That is, rules $r_t(\cdot) : \mathcal{L} \rightarrow \{0,1\}$ maps from the space of confounders to a treatment decision. Expanding on the viral load/kidney function example from earlier, consider a treatment rule that administers treatment at time $t$ only if kidney function at time $t$ is higher than some threshold $\kappa$:  $r_t(L_t) = r(L_t) = I(L_t > \kappa)$. We denote the average potential outcome under the dynamic treatment regime $a^r_{0:T} = (r(L_1), r(L_2), \dots, r(L_T) )$ as $E[Y^{a^r_{0:T}}]$. Of interest may be to compare the average difference in outcome had everyone been treated according to rule $r$ versus rule $d$: $\Psi = E[Y^{a^r_{0:T}} - Y^{a^d_{0:T}}]$

We note that these rules can be quite complex. For example, treatment at time $t$ may only be assigned if kidney function has been above $\kappa$ for the previous two periods as well as the current time period: 
$$r_t(L_{(t-2):t}) = I(L_t > \kappa_t)\cdot I(L_{t-1} > \kappa_{t-1} ) \cdot I(L_{t-2} > \kappa_{t-2})$$ 
Here, $r_t(\cdot) : \mathcal{L}^3 \rightarrow \{0,1\}$. In general, the rule may include previous treatment history as well as confounder history. In this section we consider the simple rule $r(L_t) = I(L_t > \kappa)$, but the procedure is the same for more complicated rules. 

As shown in all previous examples, Bayesian causal inference can be done quite easily provided we have posterior draws of the model parameters. Once these are obtained, computing causal contrasts is just a matter of post-processing. In this case, we only need to modify the $g$-computation post-processing steps from the previous section to sequentially set each element of the treatment vector as confounders are simulated, rather than use a pre-set treatment vector $a_{0:T}$. 

Consider the same scenario as in the static treatment setting, with posterior draw of $\omega_Y^{(m)}$ and $\omega_{L_t}^{(m)}$, but this time with a specified dynamic treatment rule $r(L_t) = I(L_t > \kappa)$. We compute a draw, $\mu^{(m)}(r)$, from the posterior of the average potential outcome under rule $r$, $E[Y^{a^r_{0:T}}]$, as follows
\begin{enumerate}
    \item Starting from $t=1$, perform the following two sub-steps sequentially until $t=T$
    \begin{enumerate}
        \item \textbf{Simulate Confounder}
        $$ \tilde L_t \sim p(L_t \mid \tilde L_{0:t-1}, \tilde a_{0:t-1}, \omega_{L_t}^{(m)})$$
        \item \textbf{Determine Treatment} according to rule
            $$  \tilde a_t = r(\tilde L_t) = I(\tilde L_t > \kappa)  $$
    \end{enumerate}
Denote these draws $\tilde L_{0:T} = ( \tilde L_0, \tilde L_1 \dots,  \tilde L_T)$ and $\tilde A_{0:T} = ( \tilde a_0, \tilde a_1, \dots, \tilde a_T)$. Repeat this a total of $B$ times to obtain $ \{  \tilde L_{0:T}^{(b)}\}_{1:B}$ and $ \{  \tilde A_{0:T}^{(b)}\}_{1:B}$
    \item \textbf{Integrate the outcome model} $\mu^{(m)}(r) = E[Y \mid a^r_{0:T}, L_{0:T} ]$ over $ \{\tilde L_{0:T}^{(b)}\}_{0:T}$ and $\{  \tilde A_{0:T}^{(b)}\}_{1:B}$ conditional on current set of draws $\omega_y^{(m)}$, under both interventions. In the viral load example, this would be 
    \begin{equation*}
        \begin{split}
            \mu^{(m)}(r) & =  \frac{1}{B}\sum_{b=1}^B E[Y \mid \tilde A_{0:T}^{(b)}, \tilde L_{0:T}^{(b)}, \omega_y^{(m)}]
        \end{split}
    \end{equation*}
\end{enumerate}
Similarly, we can draw from the posterior of average potential outcome under an alternative rule $d$, $E[Y^{a^d_{0:T}}]$. Denote this by $\mu^{(m)}(d)$. Taking the difference yields a posterior draw of $\Psi$, $\Psi^{(m)} = \mu^{(m)}(r) - \mu^{(m)}(d)$. The sum over $B$ is a Monte Carlo estimate of the integral in \eqref{eq:gcomp}. This highlights the advantage of full posterior inference. A posterior over the model parameters induces a posterior over functions of those parameters - in this case, ATEs that contrast dynamic treatment regimes.

\section{Priors over Sensitivity Parameters}\label{sc:sensitivity}
So far we have demonstrated how priors can be used to induce various correlation structures between model parameters. In Section \ref{sc:doseresponse}, we were able to estimate a smoothed causal curve by inducing correlation between neighboring points. In section \ref{sc:logistic}, we were able to estimate conditional causal contrasts for sparsely populated subgroups by shrinking their estimates towards the overall average. Lastly, in the previous section we explored ridge-like and horseshoe priors for inducing principled sparsity on a high-dimensional covariate vector. In this section, we present a different use of priors focused explicitly on causality rather than modeling - outlining how they can be used to express uncertainty about causal identification assumptions.

We consider a binary point-treatment setting with a continuous real-valued outcome. Suppose that conditional ignorability ($IA.1$) does not hold, so that $Y^a \not \perp A \mid L$, for $a\in\{0,1\}$. This implies that $E[ Y^a \mid A=1, L] \neq [ Y^a \mid A=0, L]$.  That is, the mean of each potential outcome differs between those actually treated and untreated, even after conditioning on $L$. Suppose they differ by
$$ \Delta^a(L) = E[ Y^a \mid A=1, L] - E[ Y^a \mid A=0, L] $$
This could be a result of selection bias. For instance, if higher outcome values are beneficial then $E[ Y^0 \mid A=1, L] < E[ Y^0 \mid A=0, L]$ implies those assigned to treatment would have had worse outcomes even if they had not been treated, relative to those not assigned treatment. This could be caused by ``confounding by indication'' where patients worse-off to begin with are more likely to be treated with more advanced drugs. Not accounting for this selection bias may make these drugs look ineffective and, perhaps, even harmful. 

In this setting, if we were to incorrectly assume $IA.1$, then standardization in \eqref{eq:std} would yield a biased estimated of the causal effect $\Psi = E[Y^1 - Y^0]$:
\begin{equation*}
\int_{\mathcal{L}} \{\mu(1, L) - \mu(0,L)\} dP(L)  = \Psi + \xi
\end{equation*}
where the bias term, $\xi$, is a function of $\Delta^a(L)$ and the propensity score $e(L) = P(A=1\mid L)$
\begin{equation} \label{eq:biasterm}
\xi = \int_\mathcal{L} \{\Delta^0(L) e(L) + \Delta^1(L) [1-e(L)] \}dP(L).
\end{equation}
Above, $\xi$ fully characterizes the implication of an ignorability violation on our estimate, but has a complicated form: it is a function of the treatment probability and two unknown functions, $\Delta^1(L)$ and $\Delta^0(L)$. Since ignorability is an untestable assumption, it is inherently impossible to learn about $\Delta^1(L)$ and $\Delta^0(L)$ through the observed data. To proceed, we must make assumptions about the form of the ignorability violation. The art of sensitivity analysis lies in making assumptions that balance the trade-off between the \textit{range} of violations that can be explored against the \textit{interpretability} of the sensitivity parameters. If they are not interpretable, we cannot form sensible prior beliefs about them. But if they are too simple, we will fail to explore realistic violations.

As an example, suppose that $\Delta^1(L) = \Delta^0(L) = \Delta$ so that both potential outcomes differ by some constant amount between those assigned and unassigned treatments. That is, there is some constant boost that one treatment group is getting under both hypothetical treatment interventions. We also assume this bias is constant with respect to measured covariates, so that we learn nothing about the bias by conditioning on $L$ (a worse-case scenario). In this setting, the bias reduces to $ \xi = \Delta \int_\mathcal{L} dP(L) = \Delta$. These assumptions reduce \eqref{eq:biasterm} to be a function of a single parameter which, as mentioned earlier, can be viewed as the amount of selection bias: $\Delta = E[ Y^0 \mid A=1, L] - E[ Y^0 \mid A=0, L]$. If higher $Y$ values are beneficial, then $\Delta <0$ implies treated subjects would have had outcome values $\Delta$ units lower than those not assigned treatment, even had they not been treated. This could be because of a lurking unmeasured confounder (e.g. baseline disease severity) that impacts both treatment assignment and outcome. Interpretation of magnitude will depend on the units of $Y$. If $Y$ were standardized, we could interpret $\Delta$ as a standard deviation difference in average potential outcomes between the two treatment groups. Suppose we believe that there is strong possibility of a selection bias in the $\Delta < 0$ direction and no chance of bias in the other direction, we can set $\Delta = -\Delta^*$. We could then specify a prior  $\Delta^* \sim Gam(a,b)$, which has prior mean $E[\Delta^*] = a/b$ and variance $Var[\Delta^*] = a/b^2$. For instance, if we have a prior belief of a one standard deviation bias, we can set $(a/b) = 1$ and set $b$ to, say, $b=3$. This is a fairly tight prior around $\Delta^*=1$ with standard deviation $3^{-1/2}$. 

To illustrate, we generate some synthetic data with a single binary treatment, single continuous \textit{observed} confounder, a single continuous \textit{unobserved} confounder, and a Gaussian outcome with mean being a function of treatment and both confounders. We then fit the Bayesian linear regression in Equation \eqref{eq:linearcase}, \textit{excluding} the unmeasured confounder. Appendix \ref{ap:sensitivity} describes this synthetic data generation and implementation in more detail. If we had included it, standardization would yield an accurate estimate of the ATE, $\Psi = E[Y^1 - Y^0]$, which equals $\Psi=1$ in this simulation. However, because we mistakenly exclude the unmeasured confounder, our estimate will be biased by some $\Delta$. Conducting a sensitive analysis involves specifying different priors for $\Delta$. Because we have no data about $\Delta$, the posterior is the same as the prior and so the usual standardization algorithm can be modified as follows:
\begin{enumerate}
    \item \textbf{Perform standardization} as usual to obtain $\Psi^{(m)}$. Because we are using a linear model in this simulated example, $\Psi^{(m)}$ is simply the $m^{th}$ posterior draw of the coefficient on the treatment dummy in our regression - as shown in Equation \eqref{eq:linearcase}. 
    \item \textbf{Draw sensitivity parameter} from some specified prior, e.g. $\Delta^{*(m)} \sim Gam(1,3)$, transform to get $\Delta^{(m)} = -\Delta^{*(m)}$, and compute 
    $$\Psi_{s}^{(m)} = \Psi^{(m)} - \Delta^{(m)}.$$ 
\end{enumerate}
In this case, our sensitivity analysis produces the usual posterior draws $\Psi^{(m)}$ that are perturbed by draws of $\Delta^{(m)}$. We can also view it as ``subtracting off'' the bias in \eqref{eq:biasterm} from the standardization estimate, $\Psi^{(m)}$. This perturbation incorporates our prior uncertainty regarding the magnitude of the bias due to a pre-specified form and direction of an ignorability violation. Figure \ref{fig:sensitivity} presents perturbed posteriors under three different priors for this synthetic example: $\Delta \sim N(0, sd=3^{-1/2})$, $\Delta \sim Gam(1,3)$, and $\Delta^* \sim Gam(1,3)$. Note that ignorability (i.e. no unmeasured confounding) can be expressed as a strong prior belief that $\Delta$ follows a point-mass distribution at 0, $\Delta \sim \delta_0$.  As shown in Figure \ref{fig:sensitivity}, this yields a posterior estimate centered far from $\Psi=1$. The first prior expresses symmetric belief about the direction of the bias, and so increases uncertainty in the posterior, without shifting its mean. Consequently, in Figure \ref{fig:sensitivity} we see the wider posterior interval that now has more mass around $\Psi=1$. The second prior expresses prior belief that $\Delta > 0 $ and the third expresses the belief that $\Delta<0$. Thus, the former shifts our posterior lower to correct for the upward bias and the latter shifts our posterior up to correct for the downward bias.

While sensitivity analyses around $IA$s are unique and application-specific, they follow the general procedure we outlined above:
\begin{enumerate}
    \item Find the bias induced by an $IA$ violation, $\xi$. 
    \item Make assumptions about the nature of the violation so that $\xi$ is expressed in terms of interpretable sensitivity parameters. 
    \item Express your belief about the direction and degree of the violation via priors on these sensitivity parameters. 
    \item Use draws from these priors to perturb the causal effect. 
    \item Assess the perturbed posterior.
\end{enumerate}
We contrast this approach with the usual frequentist approach that computes point and interval estimates for $\Psi$ under a pre-specified range of $\Delta$. Usually this range is wide enough so that we can see where perturbation ``reverses'' some statistically significant effect, as measured by a change in p-value from significant to non-significant. In the Bayesian approach, we see how perturbation impacts the entire posterior distribution of the estimand - telling us how posterior mean, median, quantiles, variance, etc are all affected by the uncertainty in our sensitivity parameters.

The literature on Bayesian sensitivity analysis is large and growing. For instance, McCandless et al. \citep{McCandless2007} develop a sensitivity analysis for unmeasured confounding of the effect of a binary exposure on an outcome and assess the quality of posterior inference via extensive simulations. Gustafson et al. \cite{Gustafson2010} develop a Bayesian sensitivity analysis framework for unmeasured confounding where it is assumed measured confounders are measured with error. This highlights a strength of Bayesian approach: sensitivities around multiple violations (in this case, measurement error and ignorability) can be done at once with suitable priors. Mediation analyses require more complex ignorability assumptions to estimate natural direct and indirect effects. Bayesian sensitivity analyses have been developed for such problems within the context of hazard models for survival outcomes \cite{McCandless2019}. Bayesian sensitivity analysis for mediators have also been explored with nonparametric Bayesian models \cite{Kim2017}. Other work by Gustafson et al. focus on Bayesian sensitivity for partially identified bias parameters \cite{Gustafson2014}. They discuss an application to average causal effect estimation in a randomized trial with non-compliance (i.e. not all patients randomized to treatment $A=a$ take treatment $a$).

\section{Flexible Models via Nonparametric Bayes} \label{sc:bnp}
In previous examples, we considered parametric regression models $\mu(A, L) = E[Y \mid A, L]$ that were indexed by finitely many parameters. In the Gaussian example of Section $\ref{sc:ingredients}$, the regression was determined completely by $(\theta, \beta)$. In Section \ref{sc:logistic}, the logistic regression was a function of $(\beta_w$, $\beta_v$, $\theta_0$, $\theta_{1:q})$. In our discussion of time-varying confounding, models for the confounder distribution were required at every time point, in addition to an outcome model. These models impose restrictive functional forms of the covariate and treatment effect. For instance, they assume that the treatment effects are linear and additive on some transformation of the conditional outcome mean. However, it is possible that the treatment effect is a complex, nonlinear function of $L$. Suppose all relevant confounders sufficient for $IA.1$ to hold are measured \textit{and} included in the model. Even in this scenario, msisspecification of the functional form of that model will, in general, yield inaccurate posterior causal effect estimates. In this section we will provide a brief overview of causal effect estimation using Bayesian nonparametric (BNP) models - a class of flexible models that make minimal functional form assumptions. We focus here on the point-treatment setting, with the understanding that these methods can be applied to other settings, including conditional mean modeling in $g$-computation, mediation, marginal structural models, and so on. Throughout, $D = \{Y_i, A_i, L_i \}_{1:n}$ will denote the observed data consisting of outcome, treatment, and confounder vector for $n$ independent subjects. We will define a covariate vector $X_i = (1, A_i, L_i)$ for compactness.

\subsection{Dirichlet Process Mixture Models}
We return to the linear model of Section \ref{sc:ingredients} and specify a more flexible alternative. First, define conditional regression $\mu_i(X) = X'\beta_i $. We specify the following model for the joint data distribution
\begin{equation} \label{eq:dpmix}
    \begin{split}
        Y_i \mid X_i, \beta_i, \phi_i & \sim N( \mu_i(X_i), \phi_i) \\
                    X_i \mid \theta_i & \sim  p(X_i \mid \theta_i )\\
                    \omega_i \mid G & \sim G \\
                            G  \mid \alpha, G_0 & \sim DP(\alpha G_0)
    \end{split}
\end{equation}
$\omega_i=(\theta_i, \beta_i, \phi_i)$ denotes the full parameter vector. There are two key additions in this model. First, we have saturated the model with more parameters than there are observations in the data. This is nonparametric in the sense that the number of parameters is growing with the sample size. Second, this is a \textit{generative} rather than \textit{conditional model}. That is, we model the joint distribution $p(Y_i, X_i \mid \omega_i)=p(Y_i \mid X_i,\omega_i)p(X_i \mid \omega_i)$ rather than just the conditional distribution of the outcome.

The parameters of the joint distribution follow an unknown prior, $G$. Above, we specify a Dirichlet process (DP) prior on G. Realizations of this stochastic process are \textit{discrete} random probability distributions centered around a base distribution, $G_0(\omega_i)$, with dispersion controlled by $\alpha$. This discreteness induces ties among the $\omega_i$ which, in turn, induces posterior clustering of data points. Specifically, subjects are partitioned into groups with similar joint data distributions and each group's joint is modeled using a separate $\omega_i$. In this way, the posterior conditional regression is a mixture of many cluster-specific regressions. In the machine learning literature \cite{Bishop2006} these are often called ``mixture of experts'' learners, since each component regression in the mixture (referred to as an ``expert'') has ``expertise'' in a particular region of the data. Predictions are formed by averaging over the component experts' predictions. These are distinct from ensemble models, which model the \textit{entire} data using separate candidate models - rather than assigning different data regions to different models.

\begin{figure*}[t]
    \centering
    \begin{subfigure}[b]{0.45\textwidth}
        \centering
        \includegraphics[width=\textwidth]{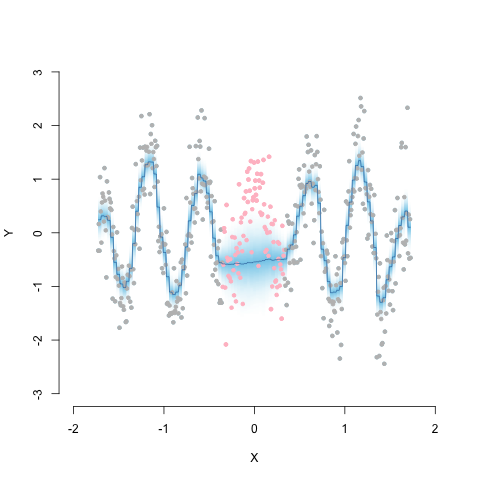}
        \caption{{ BART}}
    \end{subfigure}
    \quad
    \begin{subfigure}[b]{0.45\textwidth}  
        \centering 
        \includegraphics[width=\textwidth]{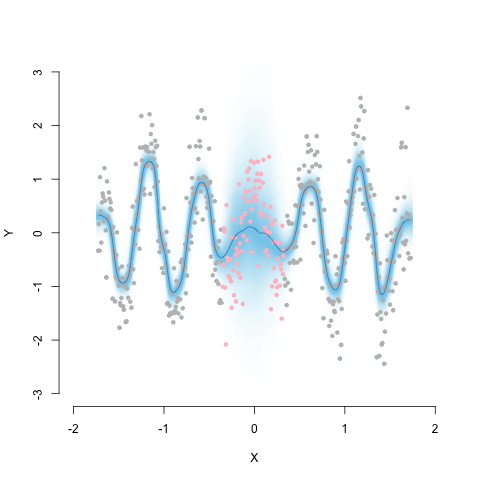}
        \caption{{ Dirichlet Process}}
    \end{subfigure}
    \vskip\baselineskip
    \begin{subfigure}[b]{0.45\textwidth}   
        \centering 
        \includegraphics[width=\textwidth]{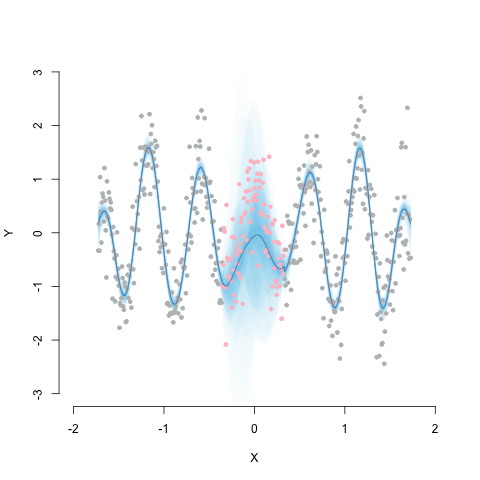}
        \caption{{ Gaussian Process}}
    \end{subfigure}
    \quad
    \begin{subfigure}[b]{0.45\textwidth}   
        \centering 
        \includegraphics[width=\textwidth]{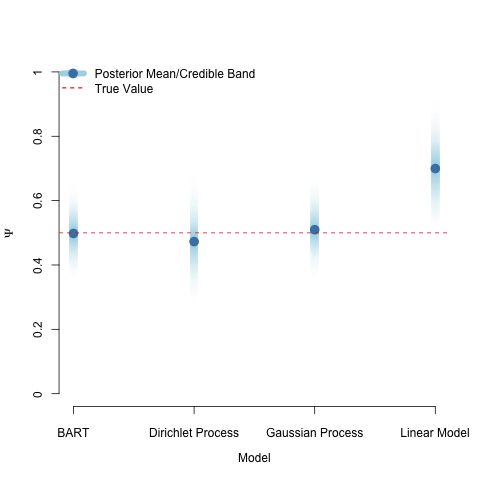}
        \caption{{ ATE Estimates}}
    \end{subfigure}
    \caption{ Training and test set predictions from three BNP models, along with ATE estimates from each. Red points indicate held-out test data. Gray points are training data. Notice for the DP and GP models, the increased uncertainty in the test region. BART, by contrast, has less uncertainty in this region. Relative to DP and GP, BART's interpolation is more rigid due to its inherent tree structure.} 
    \label{fig:predplots}
\end{figure*}

\subsection*{Induced Posterior Regression} 
Such DP mixture models have been discussed in the BNP literature for some time. Shahbaba and Neal (2007) first described a DP mixture of regressions \cite{shahbaba2007}. Blei et al.~(2011) later extended this to a DP mixture of GLMs, which generalizes \eqref{eq:dpmix} to any conditional outcome and covariate distribution in the exponential family \cite{hannah2011}. There is extensive literature on posterior sampling strategies for this model, though the most common approach in causal inference tends to be Neal's Algorithm 8 \cite{Neal2000}. We will use software to conduct the sampling, but it is instructive to show that the posterior regression can be expressed as a mixture of regressions at each iteration in the sampler. Let $\omega_{1:n}^{(m)}$ be a draw of all the subject-level parameters and let $\mu^{(m)}_i(X)=E[Y\mid X, \omega_{i}^{(m)}]$ denote the posterior regression at each iteration, given by
\begin{equation} \label{eq:dpreg}
    \begin{split}
        \mu^{(m)}_i(X) & = w_0^{(m)} \mu_0^{(m)}(X) +\sum_{i=1}^n  w_i^{(m)}\mu_i^{(m)}(X) 
    \end{split}
\end{equation}
Note that this is a mixture with $n+1$ components and mixture weights $\{w_0^{(m)},w_{1:n}^{(m)}\}$. Above, $\mu_0^{(m)}(A,L)$ is the regression under a prior draw $\beta_0^{(m)} \sim G_0$ - we will call this a ``prior regression''. The weights $w_{i}^{(m)}$ have the form
\begin{equation}
    w_i^{(m)} = \frac{ p(X\mid \theta_i^{(m)}) }{  \alpha p(X \mid \theta_0^{(m)}) + \sum_{i=1}^n p(X\mid \theta_i^{(m)}) },
\end{equation}
where $\theta_0^{(m)} \sim G_0$ is a prior draw. The weight on the prior regression is 
\begin{equation}
    w_0^{(m)} = \frac{ \alpha p(X\mid \theta_0^{(m)}) }{  \alpha p(X\mid \theta_0^{(m)}) + \sum_{i=1}^n p(X\mid \theta_i^{(m)}) }
\end{equation}
The induced posterior regression is a complex mixture of a prior regression and several subject specific regressions. Importantly, the mixture weights are covariate-dependent, allowing us to capture non-linear and non-additive effects of $X$ on the outcome. We refer to the specified distributions in \eqref{eq:dpmix} as ``local'' distributions as they are local to a particular mixture component. Even though the local model is parametric, we can approximate arbitrarily complicated distributions using a mixture of locally simple models. This is similar conceptually to approximating a complicated non-linear regression function using piecewise linear splines.

\subsection*{Local Model Choice and Hyperparameters}
Specification of the model requires specifying the local distributions. In general, model fit will not be too sensitive to these choices as the resulting regression takes a complex non-linear mixtures of these local models to fit the regression. However, desired support can be a guiding concern in making this choice. For instance, it may be desirable to choose $p(X_i \mid \theta_i)$ such that it respects the support of the elements of $X_i$. Consider a vector $X=(X_1, X_2, X_3)$ that consists of a binary, continuous/real-valued, and count confounders respectively. Assuming prior independence, we can set $p(X_i \mid \theta_i)$ to be the product of the Bernoulli, Gaussian, and Poisson distributions, with $\theta_i$ being the vector of parameters governing all three distributions. Similarly, if the outcome must be non-negative (e.g. blood pressure, cost, etc) then we could use a log-normal conditional outcome distribution instead of a Gaussian.

Just as with the local models, $G_0$ should also be set to place non-zero prior measure on the support of $\omega_i$. In model \eqref{eq:dpmix} with a single count covariate $L_i$, we could set 
$$G(\omega_i) = N_2(\beta_i; \beta^*, \Sigma^*) IG(\phi_i; a^*, b^*)Ber(p_i; p^* )IG(\lambda_i;\lambda^*)$$ 
Where, $\theta_i =(\lambda_i, p_i)$ are the parameters governing the local covariate distribution $p(X_i\mid \theta_i) = Pois(L_i; \lambda_i) Ber(A_i; p_i)$.

In the causal literature, the parameters of $G_0$ (superscripted with asterisks above) are often set using empirical Bayes principles while a relatively flat $Gamma(1,1)$ hyperprior is set on $\alpha$. Specifically, $\beta^*$ might be set to the ordinary least squares estimates, and $\Sigma^*$ may be set using the MLE covariance estimate. Empirical Bayes is a practical method of setting priors here as cross-validation would be too computationally intensive. Moreover, we typically have no substantive knowledge that could guide these choices. Centering the priors around empirical estimates also helps constrain the parameter draws to a reasonable range of the observed data. Simulation studies in a variety of scenarios show that this tends to yield adequate frequentist properties (i.e. credible intervals and point estimates with close to nominal coverage and bias, respectively, in repeated samples) \cite{Roy2017,Roy2018,Oganisian2019}. This approach is similar to Zellner's g-prior - an empirical Bayes prior popular in the Bayesian model selection literature \cite{Zellner1986}.

\subsection*{Relationship to Kernel Regression}
In this section, we discuss how the DP regression can be viewed as a Bayesian compromise between a fully empirical kernel regression and a parametric regression. 

A kernel regression estimate for a point with covariate vector $X$ is simply a weighted average of all the observed outcome values, each weighted by how ``close'' the vector $X$ is to each observed covariate. Specifically, denote the centered Gaussian kernel as $K_h(u)$ (i.e. this is the density of a Gaussian with zero mean and variance $h$). The Gaussian kernel regression \cite{Nadaraya1964} is defined as 
\begin{equation}
    \widehat E[ Y\mid X] = \sum_{i=1}^n w^{k}_i(X)\int Y \cdot K_h(Y - Y_i) dY.
\end{equation}
Note that $\int Y \cdot K_h(Y - Y_i) dY = Y_i$ is just the Gaussian mean. The weights $w^{k}_i(X)$ are given by
\begin{equation}
    w^{k}_i(X) = \frac{ K_g(X-X_i) }{\sum_i K_g(X-X_i)}.
\end{equation}
Now taking $\alpha \rightarrow 0 $ (corresponding to an improper, flat prior) in the DP regression \eqref{eq:dpreg} yields
\begin{equation}
    E[Y \mid X, \omega_{1:n}^{(m)}, D ]  = \sum_{i=1}^n  w_i^{(m)}\mu_i^{(m)}(X),
\end{equation}
with limiting weights
\begin{equation}
    w_i^{(m)} = \frac{ p(X\mid \theta_i^{(m)}) }{ \sum_{i} p(X\mid \theta_i^{(m)}) }.
\end{equation}
Comparing these equations, it is clear that the improper extreme of the DP regression becomes a type of kernel regression. In particular, if we set $p(X \mid \theta_i)$ to be Gaussian with mean $X_i$ and variance $g$ and set $\mu_i^{(m)} (X) = Y_i$, then the DP regression reduces to a kernel regression estimate. Both models are covariate-weighted mixtures of subject level conditional mean models, though the DP model is more satisfying from a statistical point of view. It outputs full posterior distribution over the regression. The kernel regression typically produces a point estimate, with uncertainty estimation being more complicated. Moreover, with the DP we can specify a covariate model, $p(X \mid \theta_i)$, that respects the support of the various covariates. This is in contrast to the kernel regression, which uses a single kernel for the whole vector.

On the other extreme, take $\alpha >> n$. Then, the DP regression becomes $ E[Y \mid X, \omega_{1:n}^{(m)}, D ] \approx w_0^{(m)} \mu_0^{(m)}(X)$. Recall here that $\mu_0$ is the regression with parameters drawn from the prior $\beta_0 \sim G_0$. The weights $w_0$ are also based on covariate parameters drawn from the prior $\theta_0 \sim G_0$. In other words, this extreme results in a completely parametric model with parameters drawn from the prior base disstribution. So we can view the DP regression as a type of posterior compromise between the kernel regression on one extreme and a parametric regression on the other. It would also be fair to say that the DP regression is a regularized version of the kernel regression. This perspective offers more insight into the role the hyperparameters of the local outcome and covariate distributions. Specifically, if $p(X \mid \theta_i)$ and $p(Y_i \mid X_i, \beta_i, \phi_i)$ are Guassian, then the variance parameters of these distribution play the same role as $h$ in the Kernel regression. Here, $h$ controls the bias-variance tradeoff. Large values of $h$ lead to a less flexible (more penalized) fit, while small values of $h$ lead to more flexible (less penalized) fit. Similarly, prior distributions on the variance parameters of these distributions that favor small values will yield a more flexible fit with less shrinkage.

\subsection*{Computing Causal Effects}
The MCMC scheme involves obtaining posterior draws of $\{\omega_{1:n}^{(m)}\}_{1:M}$, which we can use to construct the mean regression $\mu_i^{(m)}(A,L)$ at each iteration. Under $IA.1-IA.4$, we can estimate causal contrasts such as $\Psi = E[Y^1 - Y^0]$ by integrating this regression over the confounder distribution, just as in the parameter setting. Here, integration is done over a BB draw as in Section \ref{sc:bb},
\begin{enumerate}
    \item \textbf{Sample from the DP posterior} to get 
    $$ \mu^{(m)}_i(A, L) $$
    \item \textbf{Draw BB weights} 
    $$  p_{1:n}^{(m)} \mid W \sim Dir( 1_{1:n})  $$ 
    \item \textbf{Integrate} to get posterior draw of Causal Effect:
    \begin{equation*}
        \begin{split}
            \Psi^{(m)} & \approx \sum_{i=1}^n p^{(m)}_i \Big\{ \mu_i^{(m)}(1, L_i) -  \mu_i^{(m)}(0, L_i) \Big\}
        \end{split}
    \end{equation*}
\end{enumerate}
The computationally demanding portion of the above is Step 1 and can be done using off-the-shelf R packages such as \textbf{ChiRP} \cite{Oganisian2019}. This package runs the DP model in \eqref{eq:dpmix} and, by default, specifies local Gaussian distributions for non-binary and local Bernoulli distributions for binary covariates. Figure \ref{fig:predplots} visualizes predictions trained using \textbf{ChiRP}, where the conditional outcome distribution is simulated from a mixture of two damped harmonic oscillators. It also plots the ATE posterior from the DP model, computed as described above. The ATEs are computed using a synthetic data set with a binary treatment and single Gaussian confounder. In this example, the true treatment effect is a quadratic function of $L$. The figure also displays ATEs from a frequentist linear additive model, $E[Y\mid A, L] = \beta_0 + \beta_1 A + \beta_2 L$, estimated using OLS. These results are biased in this scenario. Detailed descriptions of the synthetic example used for ATE computations in Appendix \ref{ap:bnp}. This appendix also contains implementation of the ATE computation using \textbf{ChiRP}.

\subsection*{Survey of Recent DP Applications}
 The DP and related priors over random probability distributions such as the enriched DP \cite{wade2011,wade2014}, dependent DP \cite{MacEachern1999,MacEachern2000}, and centered DP \cite{Yang2010} have also been applied to causal inference. For instance, Kim et al.~(2017) employ a Dirichlet Process mixture to estimate direct and indirect effects in a mediation analysis \cite{Kim2017}. They specify a joint Gaussian model for the outcome, mediator, and confounders, and place a DP prior on the mean vector and covariance matrix. Later work applied DPs to latent mediators \cite{Kim2018}. Roy et al.~(2018) use an enriched DP to model the joint distribution of the outcome and confounders, and estimate ATEs via posterior standardization over the estimated distribution of the confounders \cite{Roy2018}. They also describe posterior imputation of missing-at-random covariates within their model. Roy et al.~(2018) use a dependent DP to estimate a marginal structural model and apply it to causal estimation with a survival outcome \cite{Roy2018}. Xu et al.~(2016)  applied a similar dependent DP model to estimate causal effects of dynamic treatment regimes \cite{Xu2016}. Xu et al.~(2018) propose an approach for estimating quantile causal effects (e.g. difference in median outcome under one intervention versus another) \cite{Xu2018}. A Bayesian Additive Regression Tree (BART) probit model is used to to model the propensity score as a function of covariates, while a Gaussian outcome model is specified conditional on the propensity score. The parameters of the joint outcome-propensity score model are given a DP mixture prior. We will describe BART models in the next section. Oganisian et al.~(2018)  specify a generative model for the joint outcome, propensity score, and confounder distribution, where the conditional outcome model is a two-part zero-inflated model \cite{Oganisian2018}. The parameters of this joint are given a DP mixture prior. Posterior standardization was conducted and a method for posterior predictive checks of positivity ($IA.4$) are proposed. Others \cite{schwartz2011} have applied DP models to adjust for post-treatment variables via principal stratification \cite{Frangakis2002}. Centered DPs have also been used to estimate heterogeneous treatment effects \cite{henderson2017}. Here, the centered DP was used as a prior for an unspecified error term distribution of an accelerated failure time model.

\subsection{Bayesian Additive Regression Trees}

The original BART approach of Chipman et al. \cite{chipman2010} models the conditional outcome distribution as a Gaussian with mean function 
\begin{equation}
    \label{eq:bart}
    \mu(X) = \sum_{j=1}^{J} g(X ; T_j, M_j) 
\end{equation}
Above, the conditional mean is modeled as a sum of predictions from $J$ regression trees, $T_j$. In this sense BART can be viewed as an ensemble learner. Specifically, $T_j$ consists of a set of nodes and splitting rules with an associated vector of terminal node parameters $M_j$. The function $g$ maps covariates $X_i$ to one of the terminal node parameters in $M_j$. The mean is then the sum of the terminal node predictions from each of the trees. The BART prior, consisting of priors on the splitting rules and terminal node parameters, is formulated to induce shrinkage towards shallow trees. This helps prevent over-fitting. This serves as a probabilistically principled alternative to pruning heuristics often used with random forests. Predictions for a toy examples are given in Figure \ref{fig:predplots}. Notice that BART produces a step function as a result of the the assumed tree structure of $\mu(X)$. This holds even as BART interpolates across the covariate space with no training data (the red points in the plot indicate held out test data).

The MCMC inference engine behind BART relies on the ``backfitting'' \cite{Breiman1985} approach, which takes posterior draws of each tree structure and their terminal node parameters sequentially. Each tree is fit using the residual from the previously fit trees as the outcome. At every iteration $m$, one such cycle through the $J$ trees yields  $T_j^{(m)}$ and $M_j^{(m)}$, which we can then use to construct a regression 
$$ \mu^{(m)}(A,L) = \sum_{j=1}^{J} g(A,L ; T^{(m)}_j, M^{(m)}_j) $$
We can use existing software in \textbf{R} such as \textbf{BayesTree} to obtain the posterior draws for $\mu^{(m)}(A, L)$ under both interventions. We first stack two test data sets $D^{a}_{test} = ( A=a, L_i)_{1:n}$ for $a\in \{0,1\}$ into a single test set $\{D^{1}_{test}, D^{0}_{test} \}$. The training data simply consists of the observed data set $D_{train} = ( Y_i, A_i, L_i)_{1:n}$. The package will then output BART estimates of $\mu(X)$ under both interventions in the stacked test set $\{ \mu^{(m)}(1, L_i), \mu^{(m)}(0, L_i) \}_{1:n}$ for $m=1,\dots M$. To compute the integral in \eqref{eq:std}, we can post-process the draws in \textbf{R} as follows. For each iteration, take a BB draw $p_{1:n}^{(m)}$ and compute 
$$ \Psi^{(m)} = \sum_{i=1}^n p_i^{(m)} \big( \mu^{(m)}(1, L_i) - \mu^{(m)}(0, L_i) \Big)$$
In this way we obtain draws from the posterior of the ATE. Our review of BART was cursory, with a focus on causal estimation. We refer the reader to Tan et al.~(2019)  for a thorough tutorial on BART and its various extensions \cite{Tan2019}. 

\subsection*{Survey of Recent BART Applications}
We now provide a brief (but by no-means exhaustive) survey of BART in interesting causal inference applications. Hill (2011) first applied BART to ATE estimation \cite{HIll2011}. BART has since enjoyed wide popularity in causal estimation. For instance, it has been used to formulate fully Bayesian semi-parametric estimation of structural mean models \cite{Zeldow2019}, fully nonparametric estimation of optimal dynamic treatment regimes \cite{Murray2018}, and estimation of causal effects in the presence of positivity violations \cite{nethery2019}. The latter augments BART with splines to extrapolate to regions of the data with deterministic treatment (i.e. non-overlap regions). Work by Hahn et al.~(2017) has focused on improving the use of BART for causal inference \cite{hahn2017}. They separate out the the treatment and confounder effects in the outcome regression, which aims to improve bias due to what the authors term ``regularization-induced confounding''. We also note that the original BART model presented here has been extended for outcomes with different support. For instance, the mean function modeled using BART can be run through a probit link when the outcome is binary. Sparapani et al. proposed using BART for survival outcomes \cite{Sparapani2016}. They use a discrete-time failure model where the probability of death at each time point is modeled with a BART probit.

\subsection{Gaussian Process (GP) Models}
Here we review another BNP approach using Gaussian process  (GP) priors for regression modeling \cite{Neal1997,Rasmussen2006}. Although less widely used in the causal literature relative to DP and BART models, GPs are popular in the BNP literature. They can be implemented in \textbf{Stan} and so may be a practical choice for applied researchers. We consider the same problem of modeling, $\mu(X)$, the mean function of a Gaussian outcome, $Y \mid X \sim N(\mu(X), \phi)$. The GP can be motivated as a prior over the space of regression functions, $\mu(X)$. We say that $\mu(X)$ follows a GP with prior mean function $\theta_0(X)$ and covariance $\bm{C} (X; \eta, \rho)$. Together with the full model, this is denoted as
\begin{equation}
\begin{split}
    Y \mid \mu(X) & \sim N\Big(\mu(X), \phi \Big) \\
            \mu(X) & \sim \mathcal{GP}\big( \theta_0, \bm{C}\big).
\end{split}
\end{equation}
Above we have suppressed dependence of $\theta_0$ and $\bm C$ on $X$ and hyperparameters $(\eta, \rho)$ for compactness. Our prior belief is that the regression function $\mu(X)$ is randomly distributed around some mean regression function $\theta_0$, with linearity and smoothness of $\mu(X)$ relative to $\theta_0$ being controlled by the hyperparameters. For example, a common prior mean function choice is $\theta_0(X) = 0$ - a hyperplane through the origin. Another, approach is to set $\theta_0(X) = X' \beta$. The latter specification centers our prior  around a linear/additive prior mean regression function, while $\eta$ and $\rho$ allow for deviations from this prior if the data are inconsistent. The covariance can have many specified forms, but we focus on the exponential-quadratic form popular in the causal literature,
\begin{equation}
    \bm C_{ij} = \eta \exp\{ -\rho|| X_i - X_j||^2 \} + .01\delta_{ij},
\end{equation}
where $||v|| = \sqrt{v'v}$ denotes the $L_2$ vector norm. $\bm C$ is the $n\times n$ matrix with elements given by $\bm C_{ij}$. Intuitively, this describes the prior belief that the regression function evaluations should be similar for two subjects with similar covariate vectors. The evaluations should differ more for two subjects who have very different covariates. The parameter $\rho$ controls how similar these function evaluations are for subjects with similar covariates. Larger $\rho$ favors more similar regression evaluations. The parameter $\eta$ controls the linearity of the regression function - with smaller $\eta$ penalizing non-linearity and \textit{a priori} favoring linear regression functions.

\textbf{Stan} can be used to sample from the posterior distribution of the regression function $\mu(X)$. Specifically, it outputs $M$ draws from the posterior of the regression function $\{\mu^{(m)}(X) \}_{1:m}$. These posterior draws are visualized in Figure \ref{fig:predplots} for both training and held-out test points. Causal ATE estimation can be done by feeding \textbf{Stan} two held-out test data sets, $D^{a}_{test} = (a, L_i)_{1:n}$ for $a\in \{0,1\}$. This returns posterior draws the regression function under both interventions $\{ \mu^{(m)}(1, L_i), \mu^{(m)}(0, L_i) \}_{1:n}$ for $m=1,\dots M$. Within \textbf{Stan}, standardization can be done using BB as described before. For each iteration, take a BB draw $p_{1:n}^{(m)}$ and compute 
$$ \Psi^{(m)} = \sum_{i=1}^n p_i^{(m)} \Big\{ \mu^{(m)}(1, L_i) - \mu^{(m)}(0, L_i) \Big\}$$
Posterior inference for the ATE using this GP model is shown in Figure \ref{fig:predplots}. Implementation details for this synthetic example are given in Appendix \ref{ap:bnp}. Finally, we note that GPs can easily accommodate outcomes with non-continuous/real support. For instance, with count outcomes we could specify $Y \mid X \sim Pois\Big( \exp( \mu(x) ) \Big)$. Here, we model $\log(E[Y\mid X]) = \mu(X)$ and place a GP prior on $\mu(X)$ as in the Gaussian case.

\subsection*{Recent Applications in Causal Inference}
Gaussian process priors have seen some usage in the causal literature. For instance, the dependent DP, used for posterior inference about marginal structural models \cite{Roy2018} and dynamic treatment regimes \cite{Xu2016} is essentially a combination of the DP and GP. Specifically, each cluster-specific regression function in the DP is assigned a GP prior. Just as the Guassian local model in \eqref{eq:dpmix} induced a posterior regression that is a mixture of linear regression functions, the dependent DP induces a posterior regression that is a mixture of GP regression functions. Other uses of GPs included modeling pollution outcomes in the presence of spatial interference (i.e. violations of $IA.3$ that exhibit spatial structure) \cite{Zigler2012} and estimation of propensity scores \cite{Vegetabile2020}.

\section{Discussion}
In this paper we reviewed causal effect estimation from a Bayesian perspective in point-treatment and time-varying treatment settings. For the latter, we outlined how to estimate causal effects of both static and dynamic treatment regimes. Both parameteric and nonparametric settings were discussed. Along the way, we discussed the utility of priors both for providing interpretable shrinkage and also for conducting causal sensitivity analyses. Throughout, we emphasize that the ad-hoc procedures we often use correspond to strongly informative priors. Throughout, we have highlighted various BNP techniques used for causal estimation in the literature. We hope that these surveys will be useful literature overviews that can serve as a starting point for those who want to delve further into these methods.

We note that our treatment of Bayesian causal estimation differs from that of Rubin \cite{Rubin1978} - which is fundamentally a finite-sample approach. In this approach, each subject's counterfactual is treated as a missing data point and the target is the posterior distribution over these missing variables, $p(\{Y_i^{1-A_i}\}_{1:n}, \mid D)$. Here, $D = \{Y^{A_i}_i , A_i, L_i \}_{1:n}$  consists of the observed potential outcomes, treatment assignment, and confounder vector. Denote the parameters governing the observed data distribution as $\omega$. By Bayes' rule we can express the desired posterior as
\begin{equation*}
    \begin{split}
        p(\{Y_i^{1-A_i}\}_{1:n}, \mid D) & =  \int p(\{Y_i^{1-A_i}\}_{1:n}, \mid \omega) p(\omega \mid D) d\omega \\
                                         & \propto  \int p(\{Y_i^{1-A_i}\}_{1:n}, \mid \omega) p(D \mid \omega) p(\omega) d\omega \\
    \end{split}
\end{equation*}
Suppose that $n_1$ of the $n$ subjects are treated. Then the likelihood is 
$$p(D \mid \omega)= p(Y_i^{1},\dots, Y_{n_1}^1, Y_{n_1+1}^0, \dots Y_{n}^{0} \mid L_{1:n}, \omega) $$
Thus this approach requires a model for the joint distribution of $p(Y^1, Y^0 \mid L, \omega)$, which is not identifiable in the data: we never observe both potential outcomes for any subject. By non-identifiable, we mean that the posterior (even if it is proper) over this joint distribution will be completely driven by the prior. This issue is not unique to Bayesian inference. For instance, the variance of the \textit{sample} average treatment effect is not identifiable from a frequentist perspective either \cite{Imbens2003}. It is a function of the covariance of the two potential outcomes, which we cannot learn. Ding et al. (2018) provide an excellent review of Bayesian causal inference from this missing data perspective \cite{Ding2018}. This missing data approach is the central idea behind the more recent PENCOMP method \cite{Zhou2019}, which uses a penalized splines to impute the missing counterfactuals. The approach described in our paper is what Ding et al.~(2018) term the ``super-population'' approach, rather than the finite-sample approach. This super-population approach focuses on estimands that are a function of the parameters governing the data generation process. Once we have a good model of the process, these estimands are simply transformations of these parameters.

\section*{Acknowledgments}
Jason Roy was supported by the National Center for Advancing Translational Sciences (NCATS), a component of the National Institute of Health (NIH) under Award Number UL1TR0030117. We thank Dr. Nandita Mitra (University of Pennsylvania) and Shira Mitchell (Civis Analytics) for very helpful comments and suggestions that improved the manuscript.

\section*{Supporting information}
All code supporting synthetic examples in this paper can be found in the accompanying GitHub repository at \url{https://github.com/stablemarkets/intro_bayesian_causal}

\bibliography{tutorial}%

\begin{thebibliography}{10}
\providecommand \doibase [0]{http://dx.doi.org/}%

\bibitem{Rubi:1974}
Rubin DB. Estimating causal effects of treatments in randomized and
  nonrandomized studies.. {\it Journal of educational Psychology} 1974\string;
  66(5)\string: 688-701.

\bibitem{Robins1986}
Robins J. A new approach to causal inference in mortality studies with a
  sustained exposure period - application to control of the healthy worker
  survivor effect. {\it Mathematical Modelling} 1986\string; 7(9)\string: 1393
  - 1512.
\newblock \href {\doibase https://doi.org/10.1016/0270-0255(86)90088-6} {doi:
  https://doi.org/10.1016/0270-0255(86)90088-6}

\bibitem{Greenland1999}
Greenland S, Pearl J, Robins JM. Causal diagrams for epidemiologic research.
  {\it Epidemiology} 1999\string: 37--48.

\bibitem{cole:fran:2009}
Cole S, Frangakis C. The consistency statement in causal inference: a
  definition or an assumption?. {\it Epidemiology} 2009\string; 20\string:
  1--5.

\bibitem{hernan2008}
Hern{\'a}n MA, Taubman SL. Does obesity shorten life? The importance of
  well-defined interventions to answer causal questions. {\it International
  Journal of Obesity} 2008\string; 32(3)\string: S8--S14.
\newblock \href {\doibase 10.1038/ijo.2008.82} {doi: 10.1038/ijo.2008.82}

\bibitem{Hudgens2008}
Hudgens MG, Halloran ME. Toward Causal Inference With Interference. {\it
  Journal of the American Statistical Association} 2008\string;
  103(482)\string: 832-842.
\newblock \href {\doibase 10.1198/016214508000000292} {doi:
  10.1198/016214508000000292}

\bibitem{imai2010}
Imai K, Keele L, Yamamoto T. Identification, Inference and Sensitivity Analysis
  for Causal Mediation Effects. {\it Statist. Sci.} 2010\string; 25(1)\string:
  51--71.
\newblock \href {\doibase 10.1214/10-STS321} {doi: 10.1214/10-STS321}

\bibitem{Daniel2013}
Daniel R, Cousens S, De~Stavola B, Kenward MG, Sterne JAC. Methods for dealing
  with time-dependent confounding. {\it Statistics in Medicine} 2013\string;
  32(9)\string: 1584-1618.
\newblock \href {\doibase 10.1002/sim.5686} {doi: 10.1002/sim.5686}

\bibitem{Baiocchi2014}
Baiocchi M, Cheng J, Small DS. Instrumental variable methods for causal
  inference. {\it Statistics in Medicine} 2014\string; 33(13)\string:
  2297-2340.
\newblock \href {\doibase 10.1002/sim.6128} {doi: 10.1002/sim.6128}

\bibitem{Lechner2010}
Lechner M. {The Estimation of Causal Effects by Difference-in-Difference
  Methods}. University of St. Gallen Department of Economics working paper
  series 2010 2010-28, Department of Economics, University of St. Gallen;
  2010.

\bibitem{Imbens2007}
Imbens G, Lemieux T. Regression Discontinuity Designs: A Guide to Practice.
  Working Paper 13039, National Bureau of Economic Research;  2007

\bibitem{andrieu2003}
Andrieu C, De~Freitas N, Doucet A, Jordan MI. An introduction to MCMC for
  machine learning. {\it Machine learning} 2003\string; 50(1-2)\string: 5--43.

\bibitem{Stan}
Carpenter B, Gelman A, Hoffman M, et al. Stan: A Probabilistic Programming
  Language. {\it Journal of Statistical Software, Articles} 2017\string;
  76(1)\string: 1--32.
\newblock \href {\doibase 10.18637/jss.v076.i01} {doi: 10.18637/jss.v076.i01}

\bibitem{gelmanbda04}
Gelman A, Carlin JB, Stern HS, Rubin DB. {\it Bayesian Data Analysis}.
\newblock Chapman and Hall/CRC.
\newblock 2nd ed.~ed. 2004.

\bibitem{Rubin1981}
Rubin DB. The {B}ayesian Bootstrap. {\it Ann. Statist.} 1981\string;
  9(1)\string: 130--134.
\newblock \href {\doibase 10.1214/aos/1176345338} {doi: 10.1214/aos/1176345338}

\bibitem{Wang2015}
Wang C, Dominici F, Parmigiani G, Zigler CM. Accounting for uncertainty in
  confounder and effect modifier selection when estimating average causal
  effects in generalized linear models. {\it Biometrics} 2015\string;
  71(3)\string: 654-665.
\newblock \href {\doibase 10.1111/biom.12315} {doi: 10.1111/biom.12315}

\bibitem{Carvalho2010}
Carvalho CM, Polson NG, Scott JG. The horseshoe estimator for sparse signals.
  {\it Biometrika} 2010\string; 97(2)\string: 465--480.

\bibitem{George1997}
George EI, McCulloch RE. APPROACHES FOR {B}ayesian VARIABLE SELECTION. {\it
  Statistica Sinica} 1997\string; 7(2)\string: 339--373.

\bibitem{McCandless2007}
McCandless LC, Gustafson P, Levy A. Bayesian sensitivity analysis for
  unmeasured confounding in observational studies. {\it Statistics in Medicine}
  2007\string; 26(11)\string: 2331-2347.
\newblock \href {\doibase 10.1002/sim.2711} {doi: 10.1002/sim.2711}

\bibitem{Gustafson2010}
Gustafson P, McCandless LC, Levy AR, Richardson S. Simplified Bayesian
  Sensitivity Analysis for Mismeasured and Unobserved Confounders. {\it
  Biometrics} 2010\string; 66(4)\string: 1129-1137.
\newblock \href {\doibase 10.1111/j.1541-0420.2009.01377.x} {doi:
  10.1111/j.1541-0420.2009.01377.x}

\bibitem{McCandless2019}
McCandless LC, Somers JM. Bayesian sensitivity analysis for unmeasured
  confounding in causal mediation analysis. {\it Statistical Methods in Medical
  Research} 2019\string; 28(2)\string: 515-531.
\newblock PMID: 28882092\href {\doibase 10.1177/0962280217729844} {doi:
  10.1177/0962280217729844}

\bibitem{Kim2017}
Kim C, Daniels MJ, Marcus BH, Roy JA. A framework for {B}ayesian nonparametric
  inference for causal effects of mediation. {\it Biometrics} 2017\string;
  73(2)\string: 401-409.
\newblock \href {\doibase 10.1111/biom.12575} {doi: 10.1111/biom.12575}

\bibitem{Gustafson2014}
Gustafson P. Bayesian inference in partially identified models: Is the shape of
  the posterior distribution useful?. {\it Electron. J. Statist.} 2014\string;
  8(1)\string: 476--496.
\newblock \href {\doibase 10.1214/14-EJS891} {doi: 10.1214/14-EJS891}

\bibitem{Bishop2006}
Bishop CM. {\it Pattern Recognition and Machine Learning (Information Science
  and Statistics)}.
\newblock Berlin, Heidelberg: Springer-Verlag .
\newblock 2006.

\bibitem{shahbaba2007}
Shahbaba B, Neal RM. Nonlinear Models Using Dirichlet Process Mixtures. 2007.

\bibitem{hannah2011}
Hannah LA, Blei DM, Powell WB. Dirichlet process mixtures of generalized linear
  models. {\it Journal of Machine Learning Research} 2011\string;
  12(Jun)\string: 1923--1953.

\bibitem{Neal2000}
Neal RM. Markov Chain Sampling Methods for Dirichlet Process Mixture Models.
  {\it Journal of Computational and Graphical Statistics} 2000\string;
  9(2)\string: 249-265.
\newblock \href {\doibase 10.1080/10618600.2000.10474879} {doi:
  10.1080/10618600.2000.10474879}

\bibitem{Roy2017}
Roy J, Lum KJ, Daniels MJ. A {B}ayesian nonparametric approach to marginal
  structural models for point treatments and a continuous or survival outcome.
  {\it Biostatistics} 2017\string; 18(1)\string: 32-47.
\newblock \href {\doibase 10.1093/biostatistics/kxw029} {doi:
  10.1093/biostatistics/kxw029}

\bibitem{Roy2018}
Roy J, Lum KJ, Zeldow B, Dworkin JD, Re VL, Daniels MJ. Bayesian nonparametric
  generative models for causal inference with missing at random covariates.
  {\it Biometrics} 2018\string; 74(4)\string: 1193--1202.
\newblock \href {\doibase 10.1111/biom.12875} {doi: 10.1111/biom.12875}

\bibitem{Oganisian2019}
Oganisian A. ChiRP: Chinese Restaurant Process Mixtures for Regression and
  Clustering. {\it Journal of Open Source Software} 2019\string; 4(35)\string:
  1287.
\newblock \href {\doibase 10.21105/joss.01287} {doi: 10.21105/joss.01287}

\bibitem{Zellner1986}
Zellner A. On assessing prior distributions and Bayesian regression analysis
  with g-prior distributions. {\it Bayesian Inference and Decision techniques}
  1986.

\bibitem{Nadaraya1964}
Nadaraya EA. On Estimating Regression. {\it Theory of Probability \& Its
  Applications} 1964\string; 9(1)\string: 141-142.
\newblock \href {\doibase 10.1137/1109020} {doi: 10.1137/1109020}

\bibitem{wade2011}
Wade S, Mongelluzzo S, Petrone S, others . An enriched conjugate prior for
  {B}ayesian nonparametric inference. {\it Bayesian Analysis} 2011\string;
  6(3)\string: 359--385.

\bibitem{wade2014}
Wade S, Dunson DB, Petrone S, Trippa L. Improving prediction from Dirichlet
  process mixtures via enrichment. {\it The Journal of Machine Learning
  Research} 2014\string; 15(1)\string: 1041--1071.

\bibitem{MacEachern1999}
MacEachern SN. Dependent nonparametric processes. {\it ASA 1999 Proceedings of
  the Section on Bayesian Statistics} 1999.

\bibitem{MacEachern2000}
MacEachern SN. Dependent Dirichlet Process.  2000.

\bibitem{Yang2010}
Yang M, Dunson DB, Baird D. Semiparametric Bayes hierarchical models with mean
  and variance constraints. {\it Computational Statistics \& Data Analysis}
  2010\string; 54(9)\string: 2172 - 2186.
\newblock \href {\doibase https://doi.org/10.1016/j.csda.2010.03.025} {doi:
  https://doi.org/10.1016/j.csda.2010.03.025}

\bibitem{Kim2018}
Kim C, Daniels M, Li Y, Milbury K, Cohen L. A {B}ayesian semiparametric latent
  variable approach to causal mediation. {\it Statistics in Medicine}
  2018\string; 37(7)\string: 1149-1161.
\newblock \href {\doibase 10.1002/sim.7572} {doi: 10.1002/sim.7572}

\bibitem{Xu2016}
Xu Y, M\"uller P, Wahed AS, Thall PF. Bayesian Nonparametric Estimation for
  Dynamic Treatment Regimes With Sequential Transition Times. {\it Journal of
  the American Statistical Association} 2016\string; 111(515)\string: 921-950.
\newblock PMID: 28018015\href {\doibase 10.1080/01621459.2015.1086353} {doi:
  10.1080/01621459.2015.1086353}

\bibitem{Xu2018}
Xu D, Daniels MJ, Winterstein AG. A {B}ayesian nonparametric approach to causal
  inference on quantiles. {\it Biometrics} 2018\string; 74(3)\string: 986-996.
\newblock \href {\doibase 10.1111/biom.12863} {doi: 10.1111/biom.12863}

\bibitem{Oganisian2018}
Oganisian A, Mitra N, Roy JA. A Bayesian nonparametric model for zero-inflated
  outcomes: Prediction, clustering, and causal estimation. {\it
  Biometrics}\string; n/a(n/a).
\newblock \href {\doibase 10.1111/biom.13244} {doi: 10.1111/biom.13244}

\bibitem{schwartz2011}
Schwartz SL, Li F, Mealli F. A {B}ayesian semiparametric approach to
  intermediate variables in causal inference. {\it Journal of the American
  Statistical Association} 2011\string; 106(496)\string: 1331--1344.

\bibitem{Frangakis2002}
Frangakis CE, Rubin DB. Principal Stratification in Causal Inference. {\it
  Biometrics} 2002\string; 58(1)\string: 21-29.
\newblock \href {\doibase 10.1111/j.0006-341X.2002.00021.x} {doi:
  10.1111/j.0006-341X.2002.00021.x}

\bibitem{henderson2017}
Henderson NC, Louis TA, Rosner GL, Varadhan R. Individualized Treatment Effects
  with Censored Data via Fully Nonparametric {B}ayesian Accelerated Failure
  Time Models. 2017.

\bibitem{chipman2010}
Chipman HA, George EI, McCulloch RE. BART: {B}ayesian additive regression
  trees. {\it Ann. Appl. Stat.} 2010\string; 4(1)\string: 266--298.
\newblock \href {\doibase 10.1214/09-AOAS285} {doi: 10.1214/09-AOAS285}

\bibitem{Breiman1985}
Breiman L, Friedman JH. Estimating Optimal Transformations for Multiple
  Regression and Correlation. {\it Journal of the American Statistical
  Association} 1985\string; 80(391)\string: 580-598.
\newblock \href {\doibase 10.1080/01621459.1985.10478157} {doi:
  10.1080/01621459.1985.10478157}

\bibitem{Tan2019}
Tan YV, Roy J. Bayesian additive regression trees and the General BART model.
  {\it Statistics in Medicine} 2019\string; 38(25)\string: 5048-5069.
\newblock \href {\doibase 10.1002/sim.8347} {doi: 10.1002/sim.8347}

\bibitem{HIll2011}
Hill JL. Bayesian Nonparametric Modeling for Causal Inference. {\it Journal of
  Computational and Graphical Statistics} 2011\string; 20(1)\string: 217-240.
\newblock \href {\doibase 10.1198/jcgs.2010.08162} {doi:
  10.1198/jcgs.2010.08162}

\bibitem{Zeldow2019}
Zeldow B, Lo~Re~III V, Roy J. A semiparametric modeling approach using
  {B}ayesian Additive Regression Trees with an application to evaluate
  heterogeneous treatment effects. {\it The annals of applied statistics.}
  2019-09\string; 13(3)\string: 1989,2010.

\bibitem{Murray2018}
Murray TA, Yuan Y, Thall PF. A {B}ayesian Machine Learning Approach for
  Optimizing Dynamic Treatment Regimes. {\it Journal of the American
  Statistical Association} 2018\string; 113(523)\string: 1255-1267.
\newblock \href {\doibase 10.1080/01621459.2017.1340887} {doi:
  10.1080/01621459.2017.1340887}

\bibitem{nethery2019}
Nethery RC, Mealli F, Dominici F. Estimating population average causal effects
  in the presence of non-overlap: The effect of natural gas compressor station
  exposure on cancer mortality. {\it Ann. Appl. Stat.} 2019\string;
  13(2)\string: 1242--1267.
\newblock \href {\doibase 10.1214/18-AOAS1231} {doi: 10.1214/18-AOAS1231}

\bibitem{hahn2017}
Hahn PR, Murray JS, Carvalho C. Bayesian regression tree models for causal
  inference: regularization, confounding, and heterogeneous effects. 2017.

\bibitem{Sparapani2016}
Sparapani RA, Logan BR, McCulloch RE, Laud PW. Nonparametric survival analysis
  using {B}ayesian Additive Regression Trees (BART). {\it Statistics in
  Medicine} 2016\string; 35(16)\string: 2741-2753.
\newblock \href {\doibase 10.1002/sim.6893} {doi: 10.1002/sim.6893}

\bibitem{Neal1997}
Neal RM. Monte Carlo implementation of Gaussian process models for {B}ayesian
  regression and classification. {\it arXiv preprint physics/9701026} 1997.

\bibitem{Rasmussen2006}
Rasmussen CE, Williams CKI. {\it Gaussian Processes for Machine Learning}.
\newblock MIT Press .
\newblock 2006.

\bibitem{Zigler2012}
Zigler CM, Dominici F, Wang Y. {Estimating causal effects of air quality
  regulations using principal stratification for spatially correlated
  multivariate intermediate outcomes}. {\it Biostatistics} 2012\string;
  13(2)\string: 289-302.
\newblock \href {\doibase 10.1093/biostatistics/kxr052} {doi:
  10.1093/biostatistics/kxr052}

\bibitem{Vegetabile2020}
Vegetabile BG, Gillen DL, Stern HS. Optimally balanced Gaussian process
  propensity scores for estimating treatment effects. {\it Journal of the Royal
  Statistical Society: Series A (Statistics in Society)} 2020\string;
  183(1)\string: 355-377.
\newblock \href {\doibase 10.1111/rssa.12502} {doi: 10.1111/rssa.12502}

\bibitem{Rubin1978}
Rubin DB. {B}ayesian Inference for Causal Effects: The Role of Randomization.
  {\it Ann. Statist.} 1978\string; 6(1)\string: 34--58.
\newblock \href {\doibase 10.1214/aos/1176344064} {doi: 10.1214/aos/1176344064}

\bibitem{Imbens2003}
Imbens GW. Nonparametric Estimation of Average Treatment Effects under
  Exogeneity: A Review. Working Paper 294, National Bureau of Economic
  Research;  2003

\bibitem{Ding2018}
Ding P, Li F. Causal Inference: A Missing Data Perspective. {\it Statist. Sci.}
  2018\string; 33(2)\string: 214--237.
\newblock \href {\doibase 10.1214/18-STS645} {doi: 10.1214/18-STS645}

\bibitem{Zhou2019}
Zhou T, Elliott MR, Little RJA. Penalized Spline of Propensity Methods for
  Treatment Comparison. {\it Journal of the American Statistical Association}
  2019\string; 114(525)\string: 1-19.
\newblock \href {\doibase 10.1080/01621459.2018.1518234} {doi:
  10.1080/01621459.2018.1518234}

\end{thebibliography}

\appendix

\section{Causal Dose Effect Example} \label{ap:dose_response}

\subsection{Data generation and implementation in Stan}

This appendix provides a more detailed walkthrough of the synthetic example and model discussed in Section \ref{sc:doseresponse}. We refer the reader to the \textbf{Stan} manual online for details about the language, syntax, and best practices. The toy example was simulated as follows. For $K=10$ doese levels $k\in \{0, \dots, 9\}$, and $n=100$ subjects, indexed by $i$ we simulate:
\begin{enumerate}
    \item single continuous confounder: 
    $$L_i \sim N(0, 1)$$
    \item treatment assignment: 
    $$A_i \mid L_i \sim P(A_i=k) \propto \expit( 1 - (2/9)\cdot k + L_i - .5kL_i ) $$
    \item outcome: 
    $$ Y_i \mid A_i, L_i \sim N( 5\cdot\Phi( A_i - 5) - 5\cdot L_i, 2 ) $$
\end{enumerate}
Above, $\Phi(\cdot)$ is the standard normal CDF. Notice that the baseline probability of treatment decreases with dose level. Reflecting a realistic scenario where fewer patients are likely to be assigned to higher doses. The confounder $L_i$ impacts both treatment and the outcome. Higher values of $L_i$ make higher dose assignments more likely (note the $-.5kL_i$ term). At the same time, higher $L_i$ lead to lower outcomes. This simulation takes place in the first several lines of dose\_response.R in the GitHub repository.

The logic behind using $\Phi$ is purely to have an interesting/realistic toy example. $A=5$ is about the middle dose level. Using $\Phi$ we are ensuring that doses much higher than the middle have diminishing returns on the outcome. Each dose increase affects the outcome less and less. Thus the true dose effect curve is $5\cdot\Phi(A - 5)$, which is plotted in red in Figure \ref{fig:doseresponse}. We need to adjust for $L_i$ because patients with higher $L$ are more likely to be treated at all levels and less likely to have higher outcomes. 

The full probability model is 
\begin{equation*}
    Y_i \mid A_i, L_i \sim N( \mu(A_i, L_i) , \phi)
\end{equation*}
Where, $\mu(A_i, L_i)$ is the conditional expectation in \eqref{eq:drmodel} - a function of $\theta_{0:K}$ and $\beta$. In the paper we discussed the priors on $\theta_{1:K}$. These took the form of a sequence of dependent Gaussian priors, as a function of $\mu$ and $\tau_k$. This likelihood is specifying in the ''model'' block of the \textbf{Stan} code DR\_model.stan:
\begin{Verbatim}[fontsize=\scriptsize]
model {
  // specify priors
  theta[1] ~ normal( 0, 10 );
  theta[2] ~ normal( 2*theta[1], 1);
  
  for(j in 3:num_A_levels){
    theta[j] ~ normal( 2*theta[j-1] - theta[j-2], 1 );
  }
  
  beta ~ normal(0, 10); 
  phi ~ cauchy(0,10);
  
  // specify likelihood 
  Y ~ normal(L*beta + A*theta , phi);
}
\end{Verbatim}
Note that in the above, an intercept is included in $L$. Notice here we have set $\mu=0$, $\tau_1=10$, and $\tau_{1:K}=1$. We specify a Gaussian prior with standard deviation (SD) 10. This is a relatively flat prior since this SD is larger than the sampling model SD$=2$. On $\phi$ we place a half-Cauchy prior with scale 10 - again, fairly flat. If we wanted to place priors on, say, $\mu$ instead of setting it at $\mu=0$, we could have instead specified
\begin{Verbatim}[fontsize=\scriptsize]
model {
  // specify priors
  mu ~ normal(0, 10);
  theta[1] ~ normal( mu, 10 );
  ...
}
\end{Verbatim}
We would also need to declare $\mu$ in the ``parameters'' block of DR\_model.stan. The same idea holds for $\tau_{0:K}$. We could also specify hyper-prior distributions for these variables. We fix these to constant values for simplicity of the illustrated examples and to maintain focus on the AR1 prior construction. Another important portion of DR\_model.stan worth highlighting is the ``generated quantities'' block. In this block, we can perform post-processing of posterior draws of parameters. For instance, we can post-process draws of $\theta_{1:K}$ to compute the curve $\Psi(k)$:
\begin{Verbatim}[fontsize=\scriptsize]
generated quantities {
  vector[num_A_levels] Psi;
  
  Psi[1] = theta[1];
  for(k in 2:num_A_levels){
    Psi[k] = theta[k] - theta[k-1];
  }
}
\end{Verbatim}
Above, we declare a vector of length num\_A\_levels (which is $K=10$ in this example). And compute $\Psi(k) = \theta_k - \theta_{k-1}$ as defined in the main text. In dose\_response.R we call the \textbf{Stan} model using the \textbf{rstan} package - which allows us to call \textbf{Stan} programs from \textbf{R}. Using the \textbf{R} function stan\_model(), we compile the Bayesian model specified in DR\_model.stan. Using the \textbf{R} function sampling() we take 500 posterior draws after a 500 draw burn-in period. Only one chain is run. In practice, more chains should be used with more posterior draws and a longer burnin period. We should check that the chains for $\Psi$ in the generated quantities block has converged. Posterior predictive checks should also be done to evaluate model fit. Guidance for convergence and posterior predictive checks is no different in this causal setting than in the general Bayesian modeling framework, so we leave details to standard Bayesian texts such as Bayesian Data Analysis \cite{gelmanbda04}.

\subsection{Implementation in SAS}

The file dose\_analysis.sas in our companion GitHub repository repeats the analysis above using PROC MCMC in \textbf{SAS}. Within PROC MCMC, we use the ``parms'' statement to declare the parameters of our model. In this case, we declare the dispersion parameter ``phi'', the nine conditional dose effects ( t1, t2, ..., t9), the confounder effect ( ``bL'' ), and the intercept (``b0'' ):
\begin{Verbatim}[fontsize=\scriptsize]
	parms phi t1-t9 bL b0;
\end{Verbatim}
The AR1 prior can be specified using the ``prior'' statement:
\begin{Verbatim}[fontsize=\scriptsize]
	prior t1 ~ normal(0, sd=10);
	prior t2 ~ normal(2*t1, sd=1);
	prior t3 ~ normal(2*t2 - t1, sd=1);
	prior t4 ~ normal(2*t3 - t2, sd=1);
	prior t5 ~ normal(2*t4 - t3, sd=1);
	prior t6 ~ normal(2*t5 - t4, sd=1);
	prior t7 ~ normal(2*t6 - t5, sd=1);
	prior t8 ~ normal(2*t7 - t6, sd=1);
	prior t9 ~ normal(2*t8 - t7, sd=1);
\end{Verbatim}

The Gaussian likelihood is specified using the ``model'' statement: 
\begin{Verbatim}[fontsize=\scriptsize]
	muA = t1*A1 + t2*A2 + t3*A3 + t4*A4 + t5*A5 + t6*A6 + t7*A7 + t8*A8 + t9*A9;
	model Y ~ normal( b0 + muA + bL*L  , sd=phi);
\end{Verbatim}
Finally, for computational efficiency, we can compute the causal effects of interest directly within PROC MCMC as simple transformations of t1, t2, ...t9 after specifying the likelihood. Note this is completely analogous to the generated quantities block in \textbf{Stan}.
\begin{Verbatim}[fontsize=\scriptsize]
	Psi[1] = t1 ;
	Psi[2] = t2 - t1;
	Psi[3] = t3 - t2;
	Psi[4] = t4 - t3;
	Psi[5] = t5 - t4;
	Psi[6] = t6 - t5;
	Psi[7] = t7 - t6;
	Psi[8] = t8 - t7;
	Psi[9] = t9 - t8;
\end{Verbatim}
 Overall, the results from \textbf{SAS} are quite similar to results in \textbf{Stan}.

\section{Conditional Causal Effects} \label{ap:partial_pool}

\subsection{Data generation and implementation in Stan}
This appendix walks through simulation and analysis of the synthetic example discussed in Section \ref{sc:logistic}, including implementation of the Bayesian bootstrap in \textbf{Stan}. The synthetic data was simulated for $n=500$ subjects (indexed here by $i$) as follows:
\begin{enumerate}
    \item Confounder $W_i \sim N(0,1)$.
    \item Stratum membership, $V_i$, with probability $P(V_i = v) = p_v$ for $v\in\{1, 2,\dots, 5\}$. Where we set 
    $$p_{1:v} = (\frac{3}{10}, \frac{3}{10}, \frac{2}{10}, \frac{1}{10}, \frac{1}{10})$$
    \item Treatment assignment as Bernoulli with probability
    $$ P(A_i \mid W_i, V_i=v) = \expit( 1 \cdot W_i + \gamma_v  ) $$
    Where $\gamma_{1:5} =(0,-.5,.5,.5,-.5)$
    \item Scalar Bernoulli outcome with probability
    \begin{equation*}
        \begin{split}
            P(Y_i \mid A_i , W_i, V_i) & = \expit[ -1 + W_i  + (1 + \sum_{v=2}^5 \eta_v I(V_i=v) ) A_i ]
        \end{split}
\end{equation*}
    Where $\eta_{2:5} = (-.5, 0, .5, .6)$ ($V=1$ is the reference).
\end{enumerate}
Again note that $W_i$ and $V_i$ both impact treatment probability and the outcome probability. The strata membership simulation mirrors practical examples where some strata are more populated (i.e. with probability $3/10$) than other strata (i.e. with probability $1/10$). In the outcome model, notice that the conditional treatment effect varies with stratum membership. Motivating the need for causal effect estimates conditional on each stratum. This simulation is done in the top portion of partial\_pool.R. 

The full probability model we specify for the outcome is that 
$$ Y_i \mid A_i, W_i, V_i \sim Ber( \mu(A_i, L_i) )$$
where $\mu(A_i, L_i)$ is the conditional expectation in \eqref{eq:logit}, reproduced here with notation specific to this example
\begin{equation*}
    \begin{split}
        E[Y_i \mid A_i, L_i] & = \sigma\Big\{\gamma + \beta_w W_i + \sum_{v=2}^5 \beta_v I(V_i=v)  + [\theta_1 + \sum_{v=2}^5 \theta_v I(V_i=v) ] A_i  \Big\}
    \end{split}
\end{equation*}
Recall here that $\sigma\{\cdot \}$ is the inverse logit link. Note here that $\theta_1$ is the conditional (on $W$) treatment effect in stratum 1 - which, in this parameterization, is the reference stratum. Similarly, $\theta_1 + \theta_v$ is the treatment effect in stratum $v$ for $v\in 2, \dots,5$.

The Bayesian model is specified in partial\_pool.stan. Below, is the ``model'' block where we specify the prior in \eqref{eq:partialpoolprior} and likelihood. In the code, ``W'' is an $n\times 1$ matrix containing each subject's confounder value and ``V'' is an $n \times 5$ matrix with $i^{th}$ row $(1, I(V_i=2), I(V_i=3), I(V_i=4), I(V_i=5) )$. Thus, $I(V_i=1)$ is the reference. The variable ``theta'' is a $5 \times 1$ vector, $\theta_{1:5}$. The variable $\beta_v$ is a $5\times 1$ vector of $V$-specific main effects, $\beta_{2:5}$ including a constant ( $\gamma$ in the model above ). 

The prior for ``theta'' is specified to induce partial pooling as discussed. We set the standard deviation of the Gaussian priors on $\theta_{1:5}$ to be $\tau=.5$. A Gaussian hyper-prior is placed on $\mu$. Note that on a logit scale this is quite flat. A Gaussian prior is also used on the coefficient of the confounder and intercept
\begin{Verbatim}[fontsize=\scriptsize]
model {
  
  // specify priors
  beta_w ~ normal(0, 1); 
  beta_v ~ normal(0, 1); 
  
  mu ~ normal(0, 1);

  theta[1] ~ normal( mu, .5 );  
  theta[2:Pv] ~ normal( mu - theta[1], .5);
  
  // specify likelihood 
  for(i in 1:N){
    Y[i] ~ bernoulli_logit( W[i]*beta_w + V[i]*beta_v + ( V[i]*theta )*A[i] ) ;
  }
}
\end{Verbatim}
As before in Appendix \ref{ap:dose_response}, we can use the generated quantities block for post-processing. As discussed in the main manuscript, this involves integrating the estimated model over a Bayesian bootstrap (BB) estimate of the conditional distribution of $W \mid V$, $P_v(W)$. Below, we include an excerpt from partial\_pool.stan that loops through each stratum of $V$ and computes a causal odds ratio for that stratum. Below, we compute the conditional mean outcome under intervention $A=1$ and $A=0$ for each subject: cond\_mean\_y1 and cond\_mean\_y0, respectively. Then we take a weighted average of these conditional means, with bootstrap weights coded as bb\_weights. Here, the \textbf{Stan} function dirichlet\_rng takes a draw from $Dir(1_{n_v})$, which is the BB posterior estimate of $P_v(W)$. This weighted average is an estimate of the \textit{marginal} mean under each intervention, coded as marg\_mean\_y1 and marg\_mean\_y0. Computing the odds ratio is done as usual using these marginal means. 

\begin{Verbatim}[fontsize=\scriptsize]
generated quantities {

  real marg_mean_y1;
  real marg_mean_y0;
  real odds_1;
  real odds_0;
  ...
  
  // cycle through strata of interest and compute causal Odds Ratio for each.
  vector[Pv] odds_ratio;

  for( v in 1:Pv ){

    // n_v = number of subjects in that stratum
    // v_start:v_end are the row indices of subjects in stratum V
    int nv = n_v[v];
    int v_start = ind[v]+1;
    int v_end = ind[v+1];

    vector[nv] cond_mean_y1;
    vector[nv] cond_mean_y0;
    vector[nv] bb_weights;

    // subset to stratum v
    matrix[nv, Pw] Wv = W[ v_start:v_end,  ];
    matrix[nv, Pv] Vv = V[ v_start:v_end,];

    // compute conditional means.
    cond_mean_y1 = inv_logit( Wv*beta_w +Vv*beta_v + Vv*theta );
    cond_mean_y0 = inv_logit( Wv*beta_w + Vv*beta_v );

    // Bayesian bootstrap weights for P_v(W)
    bb_weights = dirichlet_rng( rep_vector(1, nv) ) ;

    // taking average over bayesian bootstrap weights under both treatments
    marg_mean_y1 = bb_weights' * cond_mean_y1;
    marg_mean_y0 = bb_weights' * cond_mean_y0;

    // compute odds ratio
    odds_1 = (marg_mean_y1/(1 - marg_mean_y1));
    odds_0 = (marg_mean_y0/(1 - marg_mean_y0));
    odds_ratio[v] = odds_1/odds_0;
  }
  ...
}
\end{Verbatim}

In the sampling statement in partial\_pool.R, we run a single sampling chain consisting of 1000 posterior draws after 1000 burnin draws. The results of this computation is shown in Figure \ref{fig:partialpooling}, which is potted in partial\_pool.R.

\subsection{Implementation in SAS}
The code partial\_pool.sas in our companion GitHub repository repeats the analysis in \textbf{SAS}. The main procedures involved are PROC MCMC (as in the dose effect example) and PROC IML. Here, PROC IML (Integrated Matrix Language) is used to manipulate the posterior draws obtained from PROC MCMC and conduct the Bayesian Bootstrap. Since the PROC MCMC step is very similar to the dose effect example, here we focus on the PROC IML post-processing step. 

The first statements in PROC IML load in two datasets as matrices. First, dataset ``posterior\_draws'' is loaded and stored as matrix ``pm''. This is a matrix with each column corresponding to a parameter and each row corresponding to a posterior draw. This was an output from PROC MCMC. Second, dataset ``mcmc\_data'' is loaded and stored as matrix ``X''. This is the model matrix with $n$ rows (for each of the $n$ subjects) and covariates along the columns. Here is the relevant excerpt:

\begin{Verbatim}[fontsize=\scriptsize]
 /* Read in matrix of posterior draws from SAS to IML */
use posterior_draws; read all var _ALL_ into pm; close posterior_draws;

/* Read in model matrix from SAS to IML */
use mcmc_data; read all var _ALL_ into X; close mcmc_data;

n = nrow(X);
n_iter = nrow(pm);

/* shell to store posterior draws of Causal OR for each of the 5 strata */
OR_mat = j(n_iter, 5, 0); 
\end{Verbatim}
Next, we loop through each posterior draw, indexed by ``i'' in the code. For each draw, we loop through the strata of $V$, take a BB draw of the conditional confounder distribution $P_v(W)$, and integrate the logistic model over this BB draw of the conditional confounder distribution to attain the marginal means under each treatment. The odds ratio is computed in terms of these means. Here is the relevant excerpt:
\begin{Verbatim}[fontsize=\scriptsize]
do v = 1 to 5;
	/*X[,6] is the column containing V_i. */
	nv = sum(X[,6] = v); /* Find how many subjects in stratum v */
	idx = loc(X[,6] = v) ; /* find which obs are in stratum v */
	
	/* Draw from Dirichlet(1_{nv}) distribution  
	to do bayesian bootstrap estimate of P_v(W) */
	alpha= J(nv , 1 , 1);
	bb_w = RandDirichlet(1, alpha);
	bb_w = bb_w || 1-sum(bb_w);

	/* for each strata, compute logit of event 
	   under treatment 1 and 0: lp1, lp0*/

	/* compute reference group v=1 separately */
	if v=1 then do;
		lp1 = pm[i,2] + pm[i, 3]*X[ idx , 8] + pm[i, 8] ;
		lp0 = pm[i,2] + pm[i, 3]*X[ idx , 8] ;
	end;
			
	if v>1 then do;
		lp1 = pm[i,2] + pm[i, 3]*X[ idx , 8] + pm[i,2+v]  + (pm[i, 8] + pm[i,7+v] ) ;
		lp0 = pm[i,2] + pm[i, 3]*X[ idx , 8] + pm[i,2+v] ;
	end;
			
	/* inverse logit transform to convert to probability */
	p1 = exp(lp1)/(1+exp(lp1)); 
	p0 = exp(lp0)/(1+exp(lp0));
			
	/* bayesian bootstrap average of probability*/
	/* dot-product: bb_w is 1-X-n vector and p1, p0 are n-X-1  */ 
	mu1 = bb_w*p1; 
	mu0 = bb_w*p0;
			
	/* compute Odds Ratio for stratum v */
	OR_mat[i,v] = ( mu1/(1-mu1) ) / ( mu0/(1-mu0 ) )  ;
end;
\end{Verbatim}
Note that indexing of ``pm'' and ``X'' is so that we grab the appropriate columns of the model matrix and posterior parameters when forming the log odds ratio as a linear combination of these parameters. This is completely analogous to the ``generated quantities'' block in the \textbf{Stan} implementation. The \textbf{SAS} results are quite similar to results in \textbf{Stan}. Since we run relatively few MCMC iterations with different seeds across statistical software, some small differences are expected.

\section{Priors on Sensitivity Parameters} \label{ap:sensitivity}

\subsection{Implementation in Stan}
Here, we briefly describe using the generated quantities block in \textbf{Stan} to conduct the sensitivity analysis described in Section \ref{sc:sensitivity}. The synthetic example underlying Figure \ref{fig:sensitivity} was simulated as follows in the program sensitivity.R. For $i=1, \dots, n=100$ subjects,
\begin{enumerate}
    \item Simulate two confounders $L_i \sim N(0,1)$ and $U_i \sim N(0,1)$.
    \item Simulate treatment assignment $A_i$ from a Bernoulli with probability
    $$ P(A_i = 1 \mid L_i, U_i ) = \expit( L_i + U_i) $$
    \item Simulate outcome $Y_i$ from 
    $$ Y_i \mid A_i, L_i, U_i \sim N( A_i - L_i - 2U_i ,  1) $$
\end{enumerate}

Notice here that subjects with higher $U_i$ are more likely to be treated and have lower outcome values. Failing to adjust for $U_i$ may lead us to conclude that the treatment effect is negative, while in reality is is positive (specifically, treatment has coefficient +1 in the conditional outcome model). \\

We specify the following misspecified Bayesian model where $U_i$ is excluded:

$$ Y_i \mid L_i, A_i \sim N( \theta A_i + \beta L_i , \phi ) $$
As described in the introduction, the ATE produced by standardization from this linear conditional mean model is simply $\theta$. However, Posterior estimates of $\theta$ will be biased since we did not adjust for some unmeasured confounder $U_i$ (i.e. $IA.1$ is violated). Here, we perform the sensitivity analysis described in Section \ref{sc:sensitivity}. In the model block of sensitivity.stan, we specify the model as shown in the following excerpt
\begin{Verbatim}[fontsize=\scriptsize]
model {
  // set priors 
  theta ~ normal(0, 3);
  beta ~ normal(0, 3);
  delta1 ~ normal(0,1/sqrt(3));
  
  // specify likelihood 
  Y ~ normal(A*theta + beta*L, phi);
}
\end{Verbatim}
Notice that the sensitivity parameter here is coded as delta1 and given a standard Gaussian distribution. Now, in the generated quantities block, we compute the perturbed estimate of the ATE, coded as psi3.
\begin{Verbatim}[fontsize=\scriptsize]
generated quantities {
  ...
  real psi3;
  ...
  psi3 = theta + delta1;
}
\end{Verbatim}
This produces the posterior estimates for $\Delta \sim N(0,3^{-1/2})$ in Figure \ref{fig:sensitivity}. Note ellipses here denote ommitted code. The full code is available in the companion GitHub repository.

\subsection{Derivation of Bias}

Here we detail the derivation of the bias, $\xi$. Define the amount of bias in potential outcome $Y^a$ as 
$$ \Delta^a(L) = E[ Y^a \mid A=1,  L ]  - E[ Y^a \mid A=0,  L ]  $$
Now, take $E[Y^a]$. Iterate expectation over $L$ and then iterate once more over $A$, conditional on $L$. We get
$$ E[Y^a] = \int E[ Y^a \mid A=0, L ] + \Delta^a(L)  e(L) \ dP(L) $$
Noting that $E[Y^1 \mid A=0, L] = E[Y^1 \mid A=1, L] - \Delta^1(L)$ (by definition of $\Delta^1(L)$), we have the following expressions for each expected potential outcome
$$ E[Y^0] = \int E[ Y^0 \mid A=0, L ] + \Delta^0(L)  e(L) \ dP(L) $$
and
$$ E[Y^1] = \int E[Y^1 \mid A=1, L] - \Delta^1(L)(1 -  e(L)) \ dP(L) $$
Note that consistency allows us to drop the superscripts $a$ when conditioning on $A=a$ so that $E[Y^a \mid A=a, L] = \mu(a, L) $  . Then subtracting yields, 
$$ E[Y^1 - Y^0] = \int \mu(1, L) - \mu(0, L) dP(L)   - \Big\{ \int \Delta^1(L)(1 -  e(L)) + \Delta^0(L)  e(L) \ dP(L) \Big \}$$

Thus, under this ignorability violation, the target is equal to the usual standardization (first term on the right of the equal sign) having subtracted off the bias (the second term, which we define as $\xi$ in the main text).

\section{Time-Varying Treatments}

In the companion GitHub repository, the programs g\_comp.R and gcomp.stan simulate the synthetic example and produce the posterior inference behind Figure \ref{fig:gcomp_sparse}. Simulating and coding the analyses in this multi-time point setting is a little more tedious - involving more Stan syntax. We leave the details to the code comments. Briefly, the synthetic example contains a single binary treatment, time-varying confounder, and outcome for 10 time points. The confounder at each time point is simulated from a Gaussian with a conditional mean being a function of all previous confounder values. Treatment at each time point is simulated from a Bernoulli with probability being a function of all previous confounder and treatment values. Lastly, a single outcome at the end is simulated from a Gaussian with conditional mean being a function of all previous treatment and confounder values.

The generated quantities block demonstrates how we can simulate confounders sequentially conditional on ``always treated'' and ``never treated'' regimes. This is the type of simulation requires to compute ATEs of both static and dynamic treatment regimes outlined in the main text.

\section{Nonparametric Models} \label{ap:bnp}
This Appendix will focus on implementation details behind Section \ref{sc:bnp} - specifically the computation of ATEs in panel d of Figure \ref{fig:predplots}. We will cover implementation of DP mixtures and BART in R packages \textbf{ChiRP} and \textbf{BayesTree}, respectively, as well as GP models in \textbf{Stan} that are contained in the program npbayes\_ATE.R available in the companion GitHub repository. The synthetic data behind this example was simulated as follows. For $i=1, 2, \dots, n=500$ subjects,
\begin{enumerate}
    \item Simulate confounder $L_i \sim N(0, 1)$.
    \item Simulate treatment assignment, $A_i$, from Bernoulli with probability 
    $$ P(A_i \mid L_i ) = \expit( 1-\frac{1}{2}L_i )  $$
    \item Simulate outcome, $Y_i$, as 
    $$ Y_i \mid A_i, L_i \sim N \Big( \  (L_i + \frac{1}{2} L_i^2 ) A_i \ , \ \frac{1}{5} \Big) $$
\end{enumerate}
Note above that the conditional treatment effect is a parabolic function of $L_i$. This is a complex function form. Note that the true causal effect via standardization is:
\begin{equation*}
    \begin{split}
        \Psi & = E_L[ E[Y | A=1, L] - E[Y | A=0, L] ] \\
             & = E_L[ L_i + \frac{1}{2} L_i^2 ] \\
             & = E_L[ L_i] + \frac{1}{2} E_L[L_i^2 ] \\
             & = \frac{1}{2}
    \end{split}
\end{equation*}
The last line follows from the fact that $L$ has a standard normal distribution. A parametric model will only recover this effect if it correctly specified - a tall order for such a complex functional form. Instead, Section \ref{sc:bnp} illustrates several nonparametric approaches. 

Implementation of the model in \eqref{eq:dpmix} can be done via the fDPMix() funciton in \textbf{ChiRP}. We refer the reader to the companion web site\footnote{\url{https://stablemarkets.github.io/ChiRPsite/index.html}} and \textbf{R} help documentation for detailed information on defaults. The simulated data set is stored in an R object called d\_train in npbayes\_ATE.R. Recall that we want posterior draws of the conditional outcome mean, for each subject, under both interventions. To that end, we construct the dataset d\_test as follows:

\begin{Verbatim}[fontsize=\scriptsize]
d_a1 = data.frame(A=1, L=d_train$L)
d_a0 = data.frame(A=0, L=d_train$L)

d_test = rbind(d_a1, d_a0)
\end{Verbatim}

Note that d\_test is the observed data set stacked twice: once with treatment set to 1 for all subjects, another with treatment set to 0 for all subjects. This will allow us to obtain predictive draws for each subject under both interventions. We now feed these data sets into fDPMix() function and specify that the conditional mean outcome model to be a function of $L$ and $A$. We take 500 posterior draws after a 500 draw burnin. Initial number of clusters is set to 10. In practice, several chains with various initializations should be run and checked for mixing. 
\begin{Verbatim}[fontsize=\scriptsize]
set.seed(2)
res=fDPMix(d_train = d_train, formula = Y ~ L + A, 
           d_test = d_test, 
           iter=1000, burnin=500, init_k = 10)
\end{Verbatim}
The object res is a list containing a $2n \times$ iter-burnin matrix where the first $n$ rows are posterior predictions for each subject under treatment $A=1$ and the next $n$ rows, from $n+1$ to $2n$, are posterior predictions under treatment $A=0$. The program npbayes\_ATE.R has a short function called bayes\_boot() that performs BB standardization using these draws, as described in the test. The result is a length iter-burnin vector of posterior draws for the ATE. Additional confounders can be handled accordingly.

The other nonparametric models are implemented very similarly. In the same \textbf{R} program, we have implemented BART using the \textbf{BayesTree} package as:
\begin{Verbatim}[fontsize=\scriptsize]
bart_res = bart(x.train = d_train[, c('L','A') ], 
                y.train = d_train$Y, x.test = d_test, 
                ndpost = 500, , nskip = 500)
\end{Verbatim}
Here, we take 500 posterior draws after a 500 period burn-in. Again, we stress that in practical examples, longer burnin will likely be required. By default, this implementation runs BART using a sum of 200 trees. The function can accomodate other prior settings. We refer the reader to the \textbf{R} documentation.

The implementation of the GP regression was taken directly from the \textbf{Stan} manual section on Gaussian Processes\footnote{\url{https://mc-stan.org/docs/2_22/stan-users-guide/fit-gp-section.html}} with some minor modifications. We leave implementation details to our code and the online manual. It is quite similar to the parametric \textbf{Stan} implementations discussed in earlier appendices. \textbf{Stan} uses the following parameterization of the exponential-quadratic covariance function:
$$ \bm C_{ij} = \alpha^{2} \exp \left(-\frac{1}{2 \epsilon^{2}}||X_i - X_j||^{2}\right) $$
The parameter $\eta$ in the main manuscript corresponds to $\alpha^2$, while the parameter $\rho$ in the manuscript corresponds to $\frac{1}{2\epsilon^2}$ above. In \textbf{Stan}, we can conduct posterior inference on the hyperparameter by assigning them priors. In the program gaussian\_process\_with\_HPs\_multi.stan this is done in the model block. Here is the relevant excerpt:
\begin{Verbatim}[fontsize=\scriptsize]
model {
  ...
  rho ~ inv_gamma(5, 5);
  alpha ~ std_normal();
  ...
}
\end{Verbatim}
Here since $\alpha$ is declared to be be a non-negative parameter, the specified standard normal prior defaults to a half-normal prior.

\end{document}